\documentclass[twocolumn,english,prl,floatfix,superscriptaddress,twocolumn,amsmath,amssymb,reprint]{revtex4}

\usepackage[T1]{fontenc}
\usepackage[latin9]{inputenc}
\usepackage{color}
\definecolor{shadecolor}{rgb}{1, 0, 0}
\usepackage{verbatim}
\usepackage{float}
\usepackage{framed}
\usepackage{amsmath}
\usepackage{amssymb}
\usepackage{graphicx}

\def \W{\boldsymbol{\Omega}}

\makeatletter
\PassOptionsToPackage{caption=false}{subfig}
\usepackage{hyperref}
\hypersetup{
breaklinks=true,
colorlinks=true,
citecolor=blue,
linkcolor=blue,
filecolor=blue,
urlcolor=blue
}
\IfFileExists{lmodern.sty}{\usepackage{lmodern}}{}
\makeatother

\usepackage{babel}

\begin{document}

\title{Phonon magnetochiral effect of band-geometric origin in Weyl semimetals}
  
\author{Sanghita Sengupta}
%\email{Sanghita.Sengupta@usherbrooke.ca}
\author{M. Nabil Y. Lhachemi}
\author{Ion Garate}
\affiliation{D\'epartement de physique, Institut quantique and Regroupement Qu\'eb\'ecois sur les Mat\'eriaux de Pointe,  Universit\'{e} de Sherbrooke, Sherbrooke, Qu\'{e}bec J1K 2R1, Canada}

\date{\today}                                       
%%%%%%%%%%%%%%%%%%%%%%%%%%%%%%%%%%%%%%%%%%%%%%%%%%%%%%%%%%%%%%%%%%%%%%%%%%%%%%%

\begin{abstract}

The phonon magnetochiral effect consists of a nonreciprocity in the velocity or attenuation of acoustic waves when they propagate parallel and antiparallel to an external magnetic field.
The first experimental observation of this effect has been reported recently in a chiral magnet and ascribed to the hybridization between acoustic phonons and chiral magnons.
Here, we predict a potentially measurable phonon magnetochiral effect of electronic origin in chiral Weyl semimetals.
Caused by the Berry curvature and the orbital magnetic moment, this effect is enhanced for longitudinal phonons by the chiral anomaly.
\end{abstract}

\maketitle

%\section{Introduction}

{\em Introduction.--} In topological materials, the electronic energy bands and wave functions are characterized by nonzero integers known as topological invariants \cite{reviews}.
These invariants manifest themselves physically by virtue of peculiar electronic states localized at sample boundaries.
To date, most experimental probes of topological invariants have concentrated on electronic transport and photoemission spectroscopy.
Yet, developing alternative (possibly nonelectronic) ways to detect and exploit these invariants remains an active area of research \cite{reviews2}. 
Along this line of research, recent studies have shown that electronic topological phenomena can leave intriguing fingerprints in the properties of bulk atomic vibrations \cite{saha2015, shapourian2015, cortijo2015, andreev2016, song2016, rinkel2017, rinkel2019, chernodub2019, sukhachov2019, rajaji2019, laliberte2019}.

In the present work, we predict a new acoustic manifestation of the momentum-space geometry of electronic bands and wave functions.
We show that, in conducting crystals without inversion and mirror symmetries (chiral crystals), the electronic Berry curvature and orbital magnetic moment lead to the phonon magnetochiral effect (PMCE).
This is an effect whereby sound propagates with different speeds and attenuations in the directions parallel and antiparallel to an external magnetic field \cite{tokura2018}. 
In the bulk, PMCE is a generally weak and elusive phenomenon: thus far, it has been observed only in Cu$_2$OSeO$_3$, an insulating chiral ferrimagnet \cite{nomura2019}.
There, PMCE has been attributed to the hybridization between chiral magnons and acoustic branches of the phonon spectrum.
In contrast, the PMCE we predict takes place in non-magnetic materials with nontrivial electronic band geometry and relies on electron-phonon interactions.

For concreteness, we tailor our theory to chiral Weyl semimetals (WSM) \cite{huang2016, chang2017, tang2017, chang2018, gooth2019, rao2019, sanchez2019, schroter2019, schroter2019b, shi2019, takane2019} which, in addition to hosting intriguing transport and optical properties \cite{dejuan2017,rees2019},  possess attributes conducive to a significant PMCE.
The minimal description of a chiral WSM comprises two Weyl nodes with opposite chiralities and Berry curvatures, located at different energies.
The energy dispersion around each Weyl node is linear and the system is doped so that the Fermi surface in the absence of a magnetic field consists of two disjointed Fermi spheres of different radii. 
The phonon dispersion of chiral WSM has been theoretically studied from first principles, though only in the absence of electron-phonon interactions and magnetic fields \cite{zhang2018}.
We find that longitudinal phonons propagating along the magnetic field pump charge back and forth between Weyl nodes of opposite chirality.
The slow relaxation of this phonon-induced chiral charge imbalance amplifies the PMCE, rendering it potentially observable.
Though our calculation focuses on WSM, it can be adapted to explore the emergence of PMCE in other topologically nontrivial materials with chiral electrons, such as quantum anomalous Hall insulators \cite{liu2016}.
%, where a related PMCE may exist.

\begin{figure}[H]
  \begin{center}
\includegraphics[width=\columnwidth]{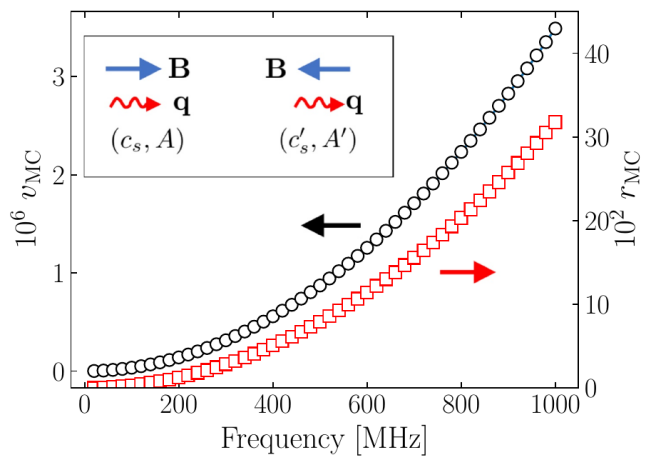}
\caption{Magnetochiral effect in the velocity ($v_{\rm MC}$) and attenuation ($r_{\rm MC}$) of longitudinal sound waves in a chiral Weyl semimetal, at fixed magnetic field and as a function of the frequency of the sound, in the diffusive regime and for modest magnetic fields (see text for details).
The inset shows the two field configurations considered (${\bf B}$ is the magnetic field and ${\bf q}$ is the phonon wave vector). In parentheses, the velocity ($c_s$) and attenuation ($A$) of the sound wave in each configuration. It follows that $v_{\rm MC}=2(c_s-c_s')/(c_s+c_s')$ and $r_{\rm MC}=2(A-A')/(A+A')$.
  %  regime ${\rm max}(\omega, v_F^2 q^2/\Gamma_A)\ll \Gamma_E\ll \Gamma_A$ and $\Gamma_A\gg v_F q$.
  %  and the parameter values are indicated in the main text.
}
 \label{fig:pmche}
  \end{center}
\end{figure}

{\em Formalism.-}
We adopt a semiclassical approach that combines the Boltzmann equation for electrons with Maxwell's equations for the electromagnetic fields generated by the lattice vibrations, and the elasticity equations for the dynamics of the lattice.
We assume that the mean free path of the electrons is long compared to the Fermi wavelength, and that the ratio between the velocity of Weyl fermions and the magnetic length is small compared to the Fermi energy. 
Although we focus on long-wavelength acoustic phonons with deformation potential coupling to electrons, other  types of electron-phonon coupling (e.g. piezoelectric) and optical phonons can likewise be incorporated in the formalism \cite{falkovsky}. 

A semiclassical theory of sound waves in metals was developed in the 1950s and 1960s, albeit for electron systems without Berry curvature \cite{kontorovich1984}.
The first attempt to augment it to topologically nontrivial materials was carried out in Ref.~[\onlinecite{andreev2016}].
The authors of this work focused on the sound attenuation in nonchiral WSM, calculated from the entropy production rate.
They compared the sound attenuation when the phonon wave vector is parallel and perpendicular to the external magnetic field, and ascribed the difference to the chiral anomaly.
The dependence of the sound attenuation was found to be an even function of the magnetic field and the phonon momentum, thereby precluding a PMCE.
Below, we obtain the full phonon dispersion (including real and imaginary parts) by solving the elasticity equations in chiral WSM, and identify new terms that are odd in both the magnetic field and the phonon momentum. 

The starting point is to calculate the distribution function $f_{\bf p}({\bf r},t)$ of electrons in a static and uniform magnetic field ${\bf B}$, in the presence of acoustic waves characterized by a displacement ${\bf u}({\bf r}, t)$ of the atomic positions with respect to equilibrium.
Here, ${\bf p}$ is the electronic momentum whereas ${\bf r}$ and $t$ are the space and time coordinates.
The function $f$ is the solution of the Boltzmann equation 
\begin{equation}
\label{eq:bol}
\left( \partial_t  +\dot{\bf r}\cdot\partial_{\bf r} + \dot{\bf p}\cdot\partial_{\bf p}\right) f_{\bf p}({\bf r},t)= I_{\rm coll}[f_{\bf p}({\bf r},t)],
\end{equation}
where $I_{\rm coll}[f]$ is the collision term to be discussed below and
\begin{align}
  \dot{\bf r} &= \partial_{\bf p}\varepsilon_{\bf p}({\bf r},t) + \dot{\bf p} \times \W_{\bf p}/\hbar\nonumber\\
  \dot{\bf p} &= e {\bf E}({\bf r},t) + e \dot{\bf r} \times {\bf B} - \partial_{\bf r}\varepsilon_{\bf p}({\bf r},t)
\end{align}
are the group velocity of an electron and the force acting on it, respectively \cite{xiao2010}.
The electron's charge is denoted as $e$ and $\W_{\bf p}$ is the Berry curvature.
The electric field ${\bf E}$ is internally produced by the lattice vibrations.
%and can be calculated by combining the Boltzmann equation and Maxwell's equations.
%The nontrivial electronic topology emerges from the Berry curvature $\W_{\bf p}$.
In addition, 
\begin{equation}
\label{eq:eps}
  \varepsilon_{\bf p}({\bf r},t) = \varepsilon_{\bf p}^{(0)} +\left(\lambda_{ij}({\bf p}) + p_i \tilde{v}_j\right) u_{i j} + ({\bf p}-m \tilde{\bf v})\cdot\dot{\bf u}
\end{equation}
is the energy of an electron in the presence of lattice vibrations.
In Eq.~(\ref{eq:eps}), $i,j\in\{x,y,z\}$, there is a sum over repeated indices,
$\varepsilon_{\bf p}^{(0)}=\varepsilon_0({\bf p}) - {\bf m}_{\bf p} \cdot {\bf B}$ is the energy of an electron in the absence of lattice vibrations, $\tilde{\bf v} = \partial_{\bf p}\varepsilon_{\bf p}^{(0)}$ is the corresponding group velocity, $\varepsilon_0({\bf p})$ is the band energy for zero magnetic field, ${\bf m}_{\bf p}$ is the orbital magnetic moment of an electron, $\lambda_{ij}({\bf p})$ is the acoustic deformation potential describing the electron-phonon coupling, and $u_{ij} = (\partial_{r_j} u_i + \partial_{r_i} u_j)/2$
%(\partial u_i/\partial r_j+\partial u_j/\partial r_i)/2$
is an element of the strain tensor.
%The geometrical 
With hindsight, we anticipate that the orbital magnetic moment will make a contribution to sound velocity and attenuation that is of the same order as that of the Berry curvature \cite{knoll2020}.

We search for a solution of Eq.~(\ref{eq:bol}) in the form 
\begin{equation}
\label{eq:f}
  f_{\bf p}({\bf r},t) = f_{\bf p}^{\rm l.e.}({\bf r},t) + \chi_{\bf p}({\bf r},t) \frac{\partial f_0(\varepsilon_{\bf p}^{(0)})}{\partial \varepsilon_{\bf p}^{(0)}},
\end{equation}
where $f_0(x)=[\exp(x)+1]^{-1}$ is the Fermi-Dirac distribution function.
%in the absence of lattice vibrations.
We limit ourselves to determining $f$ up to first order in ${\bf u}$.
In Eq.~(\ref{eq:f}), we have defined the local equilibrium distribution function 
\begin{equation}
  f_{\bf p}^{\rm l.e.}({\bf r},t) \equiv f_0 \left(\varepsilon_{\bf p}({\bf r},t) -{\bf p}\cdot\dot{\bf u}({\bf r},t) - \mu_0-\delta\mu({\bf r},t)\right),
\end{equation}
where 
$\mu_0$ is the chemical potential of the electrons in the absence of lattice vibrations, and $\delta\mu$ is the change in the chemical potential due to lattice vibrations.
Because the local equilibrium is defined in the coordinate frame that moves with the lattice, the term ${\bf p}\cdot\dot{\bf u}$ appears in it.
The second term in the right hand side of Eq.~(\ref{eq:f}) captures the phonon-induced deviations from the local Fermi Dirac distribution, which occur near the Fermi energy (we assume that the temperature of the system is low compared to the Fermi temperature).
%Accordingly, $\chi_{\bf p}({\bf r},t)$ is first-order in ${\bf u}$.

We evaluate $\delta\mu$ via the ``normalization condition'' that the total electronic density be equal to the electronic density computed from the local equilibrium distribution \cite{kontorovich1963, kontorovich1971}, i.e. $\langle\langle f({\bf r},t)\rangle\rangle = \langle\langle f_{\rm l.e.}({\bf r},t)\rangle\rangle$.
Here, the notation $\langle\langle O \rangle\rangle \equiv \int d^3 p/(2\pi\hbar)^3 O\, (1+e\W_{\bf p}\cdot{\bf B}/\hbar) $  stands for the integration of $O$ over the Brillouin zone and includes the Berry curvature correction to the density of states \cite{xiao2010}.
This condition in turn implies $\langle\chi_{\bf p}({\bf r},t)\rangle=0$, where the notation $\langle O \rangle \equiv -\langle\langle O\,\partial f_0/\partial\varepsilon^{(0)}\rangle\rangle$ stands for the Fermi surface average of $O$.

Thus far, the formalism described could be applied to an arbitrary electronic band.
For a generic system, the simplest collision term to consider in Eq.~(\ref{eq:bol}) would be \cite{kontorovich1984} 
\begin{equation}
  \label{eq:col}
I_{\rm coll}[f_{\bf p}({\bf r},t)]=  -\Gamma\chi_{\bf p}({\bf r},t) \frac{\partial f_0(\varepsilon_{\bf p}^{(0)})}{\partial\varepsilon_{\bf p}^{(0)}},
\end{equation}
$\Gamma$ being a phenomenological relaxation rate, small compared to the Fermi energy.
Hereafter, we concentrate on WSM, for which Eq.~(\ref{eq:col}) is incomplete.
For pedagogical reasons, our discussion below focuses on a minimal two-node model for  WSM.
%Yet, our main interest lies in WSM with time-reversal symmetry.
Yet, in Ref.~[\onlinecite{sm}], we provide a generalization to time-reversal-symmetric WSM with $2 n$ nodes ($n>1$), as well as to the case of multifold fermions (relevant to cubic chiral crystals such as CoSi and RhSi).
The main results derived from those more realistic models turn out to be qualitatively similar to the ones extracted from the two-node model.

In a WSM with two valleys of opposite chirality, there exist two very different relaxation rates.
First, intravalley scattering relaxes the nonequilibrium distribution function within each valley with a rate $\Gamma_A$.
Second, intervalley scattering relaxes nonequilibrium differences between the distribution functions of different valleys with a rate $\Gamma_E$.
It is commonly believed that $\Gamma_E\ll \Gamma_A$, because intervalley relaxation involves relatively large scattering wave vectors.
%By looking at our collision term, it is unclear what $\Gamma$ corresponds to.

In order to incorporate the two different relaxation rates in the problem (each of which plays a separate role in sound propagation), we use
%the following collision term for the vicinity of a valley labelled by $\alpha$:
\begin{align}
  \label{eq:col2}
  I_{\rm coll}^{(\alpha)}[f_{\bf p}^{(\alpha)}({\bf r},t)] =  -&\left[\Gamma_A \left(\chi^{(\alpha)}_{\bf p}({\bf r},t) - \frac{\langle\chi^{(\alpha)}_{\bf p}({\bf r},t)\rangle}{\langle 1^{(\alpha)}\rangle}\right)\right. \nonumber\\
    &\left. + \Gamma_E \frac{\langle\chi_{\bf p}^{(\alpha)}({\bf r},t)\rangle}{\langle 1^{(\alpha)}\rangle}\right]
  \frac{\partial f_0^{(\alpha)}(\varepsilon_{\bf p}^{(0)})}{\partial\varepsilon_{\bf p}^{(0)}},
\end{align}
where the superscript $\alpha\in\{+,-\}$ indicates that the momentum ${\bf p}$ is taken near the Weyl node $\alpha$, and $\langle 1^{(\alpha)}\rangle$ corresponds to the density of states at the (unperturbed) Fermi level on node $\alpha$.
For $\Gamma_E=\Gamma_A$, Eq.~(\ref{eq:col2}) reduces to Eq.~(\ref{eq:col}), if we project the latter to the vicinity of node $\alpha$.
For ${\bf u}=0$, Eq.~(\ref{eq:col2}) reduces to the collision term used recently in Ref. [\onlinecite{song2019}] to describe purely electronic collective modes.
%though the authors therein did not consider lattice vibrations. 
The normalization condition imposes $\sum_\alpha \langle\chi^{(\alpha)}\rangle = 0$.

In Eq.~(\ref{eq:col2}), intravalley scattering relaxes the distribution of electrons in Weyl node $\alpha$ towards a momentum-independent distribution with a local Fermi level $\mu_0+\delta\mu+\langle \chi_{\bf p}^{(\alpha)}({\bf r},t)\rangle/\langle 1^{(\alpha)}\rangle$, whereas intervalley scattering tends to equalize the electrochemical potentials at the two nodes.
If $\langle\chi_{\bf p}^{(+)}({\bf r},t)\rangle = 0$ (which implies that $\langle\chi^{(-)}_{\bf p}({\bf r},t)\rangle= 0$ through the normalization condition), the electrochemical potential is the same for the two Weyl nodes and the deviations from the Fermi-Dirac distribution happening on each node will relax through intravalley scattering alone.
If $\langle\chi_{\bf p}^{(+)}({\bf r},t)\rangle = - \langle\chi_{\bf p}^{(-)}({\bf r},t)\rangle \neq 0$, the electrochemical potential is not the same on the two nodes;
this is a manifestation of the chiral anomaly produced by lattice vibrations.
%due to electric fields generated by the sound, which (if collinear with ${|bf $}$ pumpes charge between nodes of opposite chirality. 
In that case, there will be an additional relaxation channel governed by intervalley scattering, which will tend to decrease  $|\langle\chi_{\bf p}^{(+)}({\bf r},t)\rangle|$.

To solve Eq.~(\ref{eq:bol}), we first linearize it in ${\bf u}$ and then Fourier transform it from $({\bf r},t)$ to $({\bf q},\omega)$, where $\omega$ and ${\bf q}$ are the frequency and wave vector of the lattice vibrations, respectively.
We thus obtain two equations for $\chi^{(\alpha)}_{\bf p}({\bf q},\omega)$ (one for each $\alpha$), which contain the additional unknowns ${\bf E}({\bf q},\omega)$, $\delta\mu({\bf q},\omega)$ and $\langle\chi_{\bf p}^{(\alpha)}\rangle$.
These unknowns are related to one another via the normalization condition and Maxwell's equations.
%We neglect the transverse part of the electric field.

We solve Eq.~(\ref{eq:bol}) by expanding $\chi_{\bf p}\simeq \chi_0+\chi_1$, $\delta\mu\simeq \delta\mu_0+\delta\mu_1$ and ${\bf E}\simeq {\bf E}_0+{\bf E}_1$ in powers of the magnetic field (the subscripts $0$ and $1$ stand for zeroth and first order terms in $B$, respectively).
For the purposes of PMCE, it suffices to stop the expansion at linear order.
We neglect the dependence of $\Gamma_A$, $\Gamma_E$ and $\lambda_{i j}$ in $B$.
The resulting equations for $\chi_0$ and $\chi_1$ are algebraic and can be solved analytically \cite{sm}.

The calculation is simplified in the diffusive regime with $\Gamma_A\gg q v_F^{(\alpha)}\gg \omega$ and $\Gamma_E\gg {\rm max}(\omega, (v_F^{(\alpha)})^2 q^2 /\Gamma_A)$,  which we invoke for long wavelength acoustic phonons.
Here,  $v_F^{(\alpha)}$ is the Fermi velocity at node $\alpha$.
The choice of $\Gamma_A\gg q v_F$ is motivated by the fact that the upper limit for the phonon frequency in state-of-the art ultrasound measurements is $\sim$ 1 GHz \cite{nomura2019}.
Accordingly, the shortest attainable phonon wavelength is $\sim 10 \mu{\rm m}$, much longer than the typical electronic mean free path in WSM.

%\footnote{In chiral Weyl semimetals the Fermi velocity is generally different for nodes of opposite chirality. See the Supplemental Material for a detailed treatment.}. 
Results can be further simplified by adopting the isotropic approximation for the deformation potential tensor \cite{kontorovich1984} in the vicinity of node $\alpha$,
\begin{equation}
\label{eq:lambda_iso}
  \lambda_{ij}^{(\alpha)} \simeq \lambda_1^{(\alpha)} \delta_{i j} + \lambda_2^{(\alpha)} p_i p_j/ p^2,
\end{equation}
where
${\bf p}$ is the momentun measured with respect to the node, and $\lambda^{(\alpha)}_{1 (2)}$ are constants in units of energy.
This approximation is motivated by the spherically symmetric energy dispersion around each Weyl node at $B=0$.
%If the nodes $\alpha=+$ and $\alpha=-$ were related by a crystal symmetry, we would have $\lambda_{1 (2)}^{(+)}=\lambda_{1 (2)}^{(-)}$.
%Yet, here we are interested in the situation where the two nodes are symmetry-unrelated.

Once the electronic distribution function is obtained, it is plugged into the elasticity equation describing the lattice vibrations \cite{kontorovich1984}:
\begin{equation}
  \label{eq:el}
  %\rho \ddot{u}_i = \frac{\partial\sigma_{ij}}{\partial r_j} + \left({\bf j}\times{\bf B} + {\bf F}\right)_i,
  \rho \ddot{u}_i = \partial_{r_j}\sigma_{ij} + \left({\bf j}\times{\bf B} + {\bf F}\right)_i,
\end{equation}
where $\rho$ is the mass density of the crystal, $\sigma_{ij}$ is an element of the stress tensor in the absence of conduction electrons,
\begin{equation}
  {\bf j}=-e \langle\langle f_0\rangle\rangle \dot{\bf u} + e \langle\langle \dot{\bf r} f\rangle\rangle
\end{equation}
is the total electric current (including the ionic and the electronic parts) evaluated to first order in ${\bf u}$ and to zeroth order in ${\bf B}$, and 
\begin{equation}
  % F_i=\partial \langle\langle \lambda_{i j} f\rangle\rangle/\partial r_j
   F_i=\partial_{r_j} \langle\langle \lambda_{i j} f\rangle\rangle
\end{equation}
is the $i$ component of the ``drag force'' exerted by conduction electrons on the lattice.
In Eq.~(\ref{eq:el}), we have neglected the term $(m/e) \partial_t{\bf j}$, where $m$ is the bare electron mass, because we are interested in values of the magnetic field ($\gtrsim 1 {\rm T}$) such that the free electron cyclotron frequency greatly exceeds the frequency of sound waves.

The right hand side of Eq.~(\ref{eq:el}) is (to leading order) linear in ${\bf u}$.
Thus, Eq.~(\ref{eq:el}) can be recast as an eigenvalue problem.
The corresponding eigenvectors give the polarization of the three sound waves, and the eigenvalues give their respective dispersion relations ($\omega\,\, {\rm vs}\,\, {\bf q}$).

{\em Results.-}
%Next, we summarize the main results of our calculation.
We consider an electron-doped WSM, for which $\W^{(\pm)}_{\bf p}=\pm |C| \hbar^2 {\bf p}/(2 p^3) = -{\bf m}^{(\pm)}_{\bf p} \hbar/ (e v_F^{(\pm)} p)$ ~\cite{zhong2015} and $|C|$ is the absolute value of the Chern number at a node. 
Though equal to one in our model
%(see Ref.~[\onlinecite{sm}] for generalization to more realistic models),
we keep $C$ as a bookkeeping parameter to track geometric effects in sound propagation.
We focus on the {\em change} of sound velocity and attenuation due to magnetic field, to first order in ${\bf B}$ and in the diffusive regime. 
For simplicity, we fix the phonon wave vector ${\bf q}=q_z\hat{\bf z}$ along a high symmetry direction of a cubic  crystal.
%The lenghty analytical expressions for ${\bf E}, $\delta\mu$ and $\chi$ are shown in Ref.[\onlinecite{sm}].

Let us take  ${\bf B}=B_z \hat{\bf z}$ (we will comment on the case ${\bf B}\perp{\bf q}$ briefly at the end, and more extensively in Ref.~[\onlinecite{sm}]).
%where $q_z$ and $B_z$ can be positive or negative.
Then, Eq.~(\ref{eq:el}) becomes
\begin{align}
  \label{eq:el2}
  & \rho\,\omega^2 u_i \simeq s_{i z i z} q_z^2 u_i+\frac{e}{\hbar} B_z q_z(u_i-u_z \delta_{i z})\langle\langle (\W_{\bf p}\cdot {\bf p}) f_0\rangle\rangle_0\nonumber\\
  &- i q_z \delta\mu_1 \langle\lambda_{zz}\rangle_0 \delta_{i z}+i q_z \langle\lambda_{i z}\chi_1\rangle_0,
\end{align}
where $s_{i z i z}$ is an element of the stiffness tensor and the $0$ subscript in $\langle ...\rangle_0$ and $\langle\langle ...\rangle\rangle_0$ indicates that the integrals are done at $B=0$.
The full analytical expressions for $\chi_1$ and $\delta\mu_1$ can be found in Ref.~[\onlinecite{sm}].

%$\delta\mu_1$ and $\chi_1$ are the $B$-linear parts of $\delta\mu$ and $\chi_{\bf p}$,
%The solution to Eq.~(\ref{eq:el2}) unveils a PMCE in the sound velocity and attenuation.
%$\omega=\omega_R({\bf q},{\bf B}) + i \omega_I({\bf q},

For a given acoustic mode, the PMCE in the sound velocity is defined as
\begin{equation}
  v_{\rm MC}\equiv\frac{c_s(q_z, B_z) - c_s(q_z, -B_z)}{c_s(q_z, 0)},
\end{equation}
%\begin{equation}
%  v_{\rm MC}\equiv\left[c_s(q_z, B_z) - c_s(q_z, -B_z)\right]/c_s(q_z, 0),
%\end{equation}
where
%\begin{equation}
 $ c_s({\bf q}, {\bf B})=\partial\omega_R({\bf q}, {\bf B})/\partial q$
%\end{equation}
is the sound velocity at wave vector ${\bf q}$ and $\omega_R({\bf q}, {\bf B})$ is the real part of the phonon dispersion (obtained from Eq.~(\ref{eq:el2})).
As the sound wave traverses a sample of thickness $L$, its amplitude decays by a factor  $A(q_z, B)=\exp(-\omega_I L/c_s)$ , where $\omega_I({\bf q},{\bf B})$ is the imaginary part of the phonon frequency (obtained from Eq.~(\ref{eq:el2})).
Comparing the decay factors for opposite field orientations, we define the PMCE in sound attenuation:
%as
\begin{equation}
  r_{\rm MC} \equiv \frac{A(q_z, B_z) - A(q_z, -B_z)}{A(q_z, 0)}.
%r_{\rm MC} \equiv \frac{e^{-A(q_z, B_z)} - e^{-A(q_z, -B_z)}}{e^{-A(q_z, B_z)} + e^{-A(q_z, -B_z)}},
\end{equation}
%where $A({\bf q},{\bf B})= \omega_I({\bf q},{\bf B}) L/c_s(q_z,0)$.
For longitudinal phonons (${\bf u}\parallel {\bf q}$), Eq.~(\ref{eq:el2}) yields %\cite{sm} 
\begin{align}
  \label{eq:pmce}
  v_{\rm MC} &\simeq\frac{e |C|}{\pi^2\hbar^2} \frac{q_z |q_z|B_{z}}{\rho\, c_{s}(q_z,0)}\frac{1}{\Gamma_{E}^{2}}
  \frac{\langle1^{(+)}\rangle_0 - \langle1^{(-)}\rangle_0}{\langle1^{(+)}\rangle_0+\langle1^{(-)}\rangle_0}
  \,(\delta\lambda)^2\nonumber\\
  r_{\rm MC} &\simeq -\frac{7 e |C|}{12\pi^{2}\hbar^{2}}\frac{q_{z} |q_z| B_{z} L}{\rho\, c_s(q_z,0)^2}\frac{1}{\Gamma_{E}}\frac{\langle\lambda_{zz}\rangle_0}{\langle1\rangle_0}
  \,\delta\lambda,
\end{align}
%\begin{widetext}
%\begin{align}
%  \label{eq:pmce}
%  v_{\rm MC} &\simeq\frac{e |C|}{\pi^2\hbar^2} \frac{q_z |q_z|B_{z}}{\rho\, c_{s}(q_z,0)}\frac{1}{\Gamma_{E}^{2}}
%  \frac{\langle1^{(+)}\rangle_0 - \langle1^{(-)}\rangle_0}{\langle1^{(+)}\rangle_0+\langle1^{(-)}\rangle_0}
%  \bigg(\lambda_1^{(+)}-\lambda_1^{(-)} + \frac{\lambda_2^{(+)}}{3}-\frac{\lambda_2^{(-)}}{3}\bigg)^{2}\nonumber\\
%  r_{\rm MC} &\simeq -\frac{7 e |C|}{12\pi^{2}\hbar^{2}}\frac{q_{z} |q_z| B_{z} L}{\rho\, c_s(q_z,0)^2}\frac{1}{\Gamma_{E}}\fra%c{\langle\lambda_{zz}\rangle_0}{\langle1\rangle_0}
%  \bigg(\lambda_1^{(+)}-\lambda_1^{(-)} + \frac{\lambda_2^{(+)}}{3}-\frac{\lambda_2^{(-)}}{3}\bigg),
%\end{align}
%\end{widetext}
where $\delta\lambda \equiv \lambda_1^{(+)} - \lambda_1^{(-)} + \lambda_2^{(+)}/3-\lambda_2^{(-)}/2$ and  we have omitted terms that are smaller by at least a factor $\Gamma_A/\Gamma_E$; some of these terms are intrinsic (independent of $\Gamma_A$ and $\Gamma_E$) but quantitatively negligible.
When the deformation potentials and the Fermi level density of states differ strongly between the two nodes of opposite chirality, we have $r_{\rm MC}/v_{\rm MC}\sim \Gamma_E L/c_s(q_z,0)\gg 1$.
For long wavelengths, no significant error is made by replacing  $c_s(q_z,0) \to c_s(0,0)\equiv c_s(0)$ in Eq. (\ref{eq:pmce}).
For a $2n$-node WSM model ($n>1$), time-reversed partners make additive contributions to the PMCE and, as a result, $v_{MC}$ and $r_{MC}$ become larger \cite{sm}.
All else unchanged, PMCE is likewise enhanced when the bands at the Fermi surface have larger Chern numbers, which is the case in chiral WSM like CoSi or RhSi \cite{sm}.
%whose bands are endowed with larger Chern numbers.
% Chern numbers in this material are conducive to a stronger PMCE. 
%Yet, provided that their acoustic  deformations are of the order of 1 eV.
% That said, we do not know the values of the deformation potential in these materials, %and the effect of the Fermi surface anisotropy in PMCE needs to be investigated.

% enhanced by a factor of two
%$\lambda_{1(2)}^{(+)}+\lambda_{1(2)}^{(-)}\simeq |\lambda_{1(2)}^{(+)}-\lambda_{1(2)}^{(-)}|$ and $\langle1^{(+)}\rangle_0 + \langle1^{(-)}\rangle_0\simeq |\langle1^{(+)}\rangle_0 - \langle1^{(-)}\rangle_0|$)

%The fact that $r_{\rm MC}\gg v_{\rm MC}$ does not however necessarily mean that the PMCE is more easily observable in sound attenuation, because $v_{\rm MC}$ can be measured with much higher precision than $r_{\rm MC}$.

Equation (\ref{eq:pmce}) is the central result of this work.
As expected for a magnetochiral effect, $v_{\rm MC}$ and $r_{\rm MC}$ are odd in $q_z$ and $B_z$.
Interestingly, they are also proportional to $|C|$, which means that they originate entirely from the momentum-space geometry of electronic Bloch wave functions.
The underlying intuition is that the electronic chirality of WSM, encoded in $C$, is communicated to sound propagation via the electron-phonon coupling.

Because the net chirality of a WSM is zero (Nielsen-Ninomiya (NN) theorem \cite{nielsen1981}), it takes a special type of electron-phonon interaction to transfer the chirality of individual Weyl fermions to the sound.
%One interesting aspect of $v_{\rm MC}$ and $r_{\rm MC}$ (which also applies to the omitted terms) is that they are proportional to the Chern number 
Indeed, as a result of the NN theorem, $v_{\rm MC}$ and $r_{\rm MC}$ vanish when the deformation potential is identical in nodes of opposite chirality ($\delta\lambda=0$).
This can also be understood on the basis of symmetry.
The PMCE of Eq.~(\ref{eq:pmce}) originates from terms in the sound dispersion that contain ${\bf q}\cdot{\bf B}$.
The sound frequency must be a scalar \cite{scalar}, while ${\bf q}\cdot {\bf B}$ is a pseudoscalar.
Thus, terms containing ${\bf q}\cdot{\bf B}$ need to have pseudoscalar proportionality factors.
Such factors emerge naturally in the presence of {\em pseudoscalar phonons}, whose deformation potentials have the same magnitude and opposite sign on Weyl nodes of opposite chirality related by an improper symmetry (hence $\delta\lambda\neq 0$).

In principle, there exist nonchiral crystals that host pseudoscalar acoustic phonons; they belong to the point groups
\{$C_{1h} (A'')$, $C_{2h} (A_u)$, $C_{3h} (A'')$ and $S_4 (B)$\} (noncentrosymmetric) and \{$C_i (A_u)$, $C_{4h} (A_u)$, $C_{3i} (A_u)$, $C_{6h} (A_u)$\} (centrosymmetric), where the letters in parentheses denote the irreducible representation of the pseudoscalar acoustic phonon.
To get to this result, we have identified the pseudoscalar irreducible representations in all point groups \cite{anastassakis1972}
and have checked whether they overlap with the vector representations (indeed, long-wavelength acoustic phonons transform according to the vector representation because $q=0$ acoustic phonons are pure translations).

To our knowledge, there are no examples of WSM belonging to the aforementioned point groups.
This is where chiral crystals like CoSi become important: because these crystals do not possess any improper symmetry operations, Weyl nodes of opposite chirality are not related by any symmetry and thus they will be subjected to unequal deformation potentials ($\delta\lambda\neq 0$).
Then, $\delta\lambda$ furnishes the pseudoscalar quantity that is needed in order to realize the PMCE.
Another pseudoscalar quantity that appears in the theory is the difference in the Fermi-level density of states between two nodes of opposite chirality, which is nonzero in chiral crystals.

Another point to highlight in Eq.~(\ref{eq:pmce}) is that $v_{\rm MC}$ and $r_{\rm MC}$ scale as $1/\Gamma_E^2$ and $1/\Gamma_E$, respectively, when ${\rm max}(\omega, v_F^2 q^2/\Gamma_A)\ll \Gamma_E\ll \Gamma_A$.
The mechanism underlying this dependence is that longitudinal phonons propagating collinearly with the magnetic field generate a dynamical chiral population imbalance, whose magnitude is set by the intervalley relaxation rate \cite{similarities}.
Because this relaxation rate is slow, the PMCE is enhanced.
%This dependence on the intervalley relaxation rate connects the PMCE to the dynamical chiral population imbalance generated by lattice vibrations through the chiral anomaly.

One last important characteristic of Eq.~(\ref{eq:pmce}) is that $v_{\rm{MC}}$ and $r_{\rm{MC}}$ are not negligible.
Figure \ref{fig:pmche} displays Eq.~(\ref{eq:pmce}) as a function of $c_s(0) q$.
For reasonable parameter values ($B= 1\, {\rm T}$, $\hbar\Gamma_E=0.01\, {\rm meV}$, $\lambda_{1 (2)}^{(+)} = 2.0\, {\rm eV}$, $\lambda_{1 (2)}^{(-)} = 1.0\, {\rm eV}$, $\rho=10^4\, {\rm kg/m}^{3}$, $c_s(0)=2\times10^3\, {\rm m/s}$, $L=1 {\rm cm}$), $v_{\rm MC}$ and $r_{\rm MC}$ exceed the threshold of detectability (which is $\simeq 10^{-6}$ for $v_{\rm MC}$ \cite{nomura2019}, and $\simeq 10^{-3}-10^{-2}$ for $r_{\rm MC}$ \cite{jeff}).
%and  and a wave vector of $q=0.5\times 10^6\, {\rm m}^{-1}$, we get $v_{\rm MC}\simeq 3\times 10^{-6}$.
%This can be detected in state-of-the-art ultrasound velocity measurements, whose resolution is $\simeq 10^{-6}$ \cite{nomura2019}.
%For the same parameters and $L=1\, {\rm cm}$, we obtain values of $r_{\rm MC}$ well above the experimental resolution of $\simeq 10^{-3}-10^{-2}$.
Clearly, the observability of $v_{\rm MC}$ and $r_{\rm MC}$  is aided by the slowness of the intervalley relaxation time.
The value chosen here ($1/\Gamma_E\simeq 50$ ps) is within the range discussed in the literature \cite{parameswaran2014, zhang2015, mehdi2019}.
%Another key parameter determining the observability of PMCE is the acoustic deformation potential; though we have adopted a typical value, the actual number in existing chiral WSM is not known.
%value adopted here for the deformation potentials is likewise standard, though the numbers for materials like CoSi and RhSi are not known.

Finally, we remark that the situation is quite different for transverse phonons (${\bf q}\cdot{\bf u}=0$).
%We emphasize that this result holds only for longitudinal phonons with ${\bf q}\cdot{\bf B}\neq 0$.
In our approximation, these phonons do not generate a chiral population imbalance \cite{sm}.
%This statement agrees with that found using a different approach in Ref.~[\onlinecite{spivak2016}].
Accordingly, their PMCE is much weaker (by at least a factor $\Gamma_A/\Gamma_E$).
This statement also applies to all three phonon modes in the configuration ${\bf q}\cdot{\bf B}=0$.

%The strong quantitative dependence of the PMCE in the polarization of the sound waves can be a diagnostic tool for the experimental detection of the effect.
%Another way to distinguish the PMCE experimentally is through its field-dependence (linear in the semiclassical regime) and frequency-dependence (quadratic at frequencies lower than the intervalley scattering rate).

%{\bf Check?}
%Finally, we comment on the configuration ${\bf q}\cdot{\bf B}=0$.
%In this case
%Thus far, we have discussed the case with ${\bf q}\cdot{\bf B}\ne0$.
%One can readilly repeat the analysis  
%It turns out that in this case there is no dependence of the sound dispersion in magnetic field to first order.

{\em Conclusions.-}
We have theoretically predicted a phonon magnetochiral effect of band-geometric origin in chiral Weyl semimetals.
%Provided that the acoustic deformation potentials in these materials are of the order of an eV, the predicted effect may be be observable.
Analogous effects may occur for other quasiparticles (such as photons and magnons).
%in other topological materials, and for different topological invariants.
In order to experimentally detect the predicted effect and to rule out background signals, one should 
%(i) carefully calibrate the zero of the magnetic field (an offset in magnetic field would result in a spurious PMCE),
(i)  make sure that the PMCE signal changes sign when (and only when) the relative direction between the magnetic field and the phonon wave vector is reversed, (ii) confirm that a reflection measurement (in which sound waves travel an equal distance in the direction parallel and antiparallel to the field) does not result in a PMCE signal, (iii) observe an increase in the signal with the magnetic field (linearly, in the semiclassical regime) and with the frequency of the sound (quadratically, when the frequency is smaller than the intervalley scattering rate).

%for longitudinal phonons by the slow relaxation rate of the dynamical valley imbalance, induced by the lattice vibrations through the chiral anomaly. 
%The values of acoustic deformation potentials
%It will be important to determine the deformation potentials and the intervalley scattering rates in CoSi, in view of an experimental discovery of a PMCE of topological origin.

%The success of such attempts may hinge on the magnitudes of the acoustic deformation %potential in this material, which are not known to us. That said, the larger Chern %numbers in these materials are promising in that they lead to a stronger PMCE. 

%Future avenues of research include adapting our semiclassical theory to other topological materials and carrying out a fully quantum mechanical calculation in strong magnetic fields. 
%It will be of interest to adapt 
%systems, where the phonon magnetochiral effect interplays with nontrivial band topology.
%It will likewise be useful to perform a fully quantum mechanical calculation in the strong magnetic field regime.

{\em Acknowledgements.-}
This research has been financed in part by the Canada First Research Excellence Fund, the Natural Science and Engineering Council of Canada, the Fonds de Recherche du Qu\'ebec Nature et Technologies, and by the National Science Foundation under Grant No. NSF PHY-1748958.
I.G. acknowledges the hospitality of the Kavli Institute for Theoretical Physics, where this work was finalized. 
We are grateful to O. Antebi, D. Pesin, J. Quilliam and M. A. H. Vozmediano for informative discussions.
We thank C. Ethier for her technical assistance in the early stages of this work.

%reference file%
\input{refs1.tex}
% supplementary material
%-----------------------------------
%%%%%%%%%% Prefix a "S" to all counters and reset the counter %%%%%%%%%%
\setcounter{equation}{0}
\setcounter{figure}{0}
\setcounter{table}{0}
\makeatletter
\renewcommand{\theequation}{S\arabic{equation}}
\renewcommand{\thefigure}{S\arabic{figure}}
\renewcommand{\thetable}{S\arabic{table}}
\renewcommand{\bibnumfmt}[1]{[S#1]}
\renewcommand{\citenumfont}[1]{S#1}
%%%%%%%%%% Prefix a "S" to all counters and reset the counter %%%%%%%%%%

% BEGIN SUPPLEMENTARY MATERIAL 
%%%%%%%%%% Merge with supplemental materials %%%%%%%%%%
\clearpage
\pagebreak
\setcounter{page}{1}
\onecolumngrid
\widetext
\begin{center}
\textbf{{\large Supplemental material for ``Phonon magnetochiral effect of band-geometric origin in Weyl semimetals"}}\\[1em]

\text{Sanghita Sengupta, M. Nabil Y. Lhachemi, and Ion Garate}
\vspace{0.2 cm}

\text{\em D\'epartement de physique, Institut Quantique and Regroupement Qu\'eb\'ecois sur les Mat\'eriaux de Pointe}
\text{\em Universit\'e de Sherbrooke, Sherbrooke, Qu\'ebec J1K 2R1, Canada}
\end{center}

\noindent
In this supplemental material (SM) we provide additional information and details about the formalism described in the main text.
The outline of the SM is as follows:
\vspace{0.2 cm}

I. Equations of motion of an electron in the presence of lattice vibrations.

II. Boltzmann kinetic equation (BKE).

III. Solution of the BKE to zeroth order in the magnetic field.

IV. Solution of the BKE to first order in the magnetic field.

V. Elasticity equations for lattice vibrations in the presence of Weyl fermions.

VI. Velocity and attenuation of sound waves in a two-node model of Weyl semimetals: phonon magnetochiral effect.

VII. Phonon magnetochiral effect in a Weyl semimetal model of 2$n$-nodes ($n>1$).

VIII. Phonon magnetochiral effect in a model of multifold chiral fermions.

%We give explicit calculations for (i) equations of motion of the electron in presence of lattice vibrations, (ii) we then use the electronic equations of motion to derive the Boltzmann kinetic equation for our model, (iii) solution of the Boltzmann kinetic equation (BKE) derived perturbatively in magnetic field, (iv) elasticity equations for the lattice vibrations written with the solution of the BKE and (v) velocity and attenuation of sound propagation characterizing the phonon magnetochiral effect.

\subsection*{I. Equations of motion of an electron in presence of lattice vibrations}
\label{sec:motion}
The semiclassical equations of motion for an electron at position ${\bf r}$ with momentum ${\bf p}$ are \cite{xiao2010}
\begin{align}
  \label{r}
  \dot{\bf{r}} & = \partial_{\bf{p}}\varepsilon_{\bf{p}}({\bf{r}},t) +\dot{\bf{p}} \times \boldsymbol{\Omega}_{\bf p}/\hbar\\
%\end{equation}
%\begin{equation}\label{p}
\label{p}
  \dot{\bf{p}} &= e{\bf E}({\bf{r}},t) + e \dot{{\bf r}} \times {\bf B} - \partial_{\bf{r}}\varepsilon_{\bf{p}}({\bf{r}},t),
\end{align}
where $\boldsymbol{\Omega}_{\bf p}$ is the Berry curvature,
%The dispersion relation in presence of orbital magnetic moment and presence of phonons is given as
\begin{equation}
\varepsilon_{\bf p}({\bf r},t) = \varepsilon_{\bf p}^{(0)} + \delta{\varepsilon}({\bf r},t)
\end{equation}
is the energy of the electron in the presence of lattice vibrations and magnetic fields, 
\begin{equation}
    \varepsilon_{\bf p}^{(0)}(\textbf{p}) = \varepsilon_0(\textbf{p}) -  \textbf{m}_{\bf p}\cdot\textbf{B}
\end{equation}
is the energy in the absence of lattice vibrations, $\varepsilon_0$ is the band energy in absence of lattice vibrations and magnetic fields, $\textbf{m}_{\bf p}$ is the orbital magnetic moment and 
\begin{equation}
  \label{eq:de}
    \delta\varepsilon_{\bf p}({\bf r},t) = \left(\lambda_{ij}({\bf p}) + p_{i}\Tilde{v}_{j}\right) u_{ij}({\bf r},t) + p_{i}\dot{u_{i}} - m\Tilde{v}_{i}\dot{u}_{i}
\end{equation}
is the contribution of lattice vibrations to the electron's energy (written in the lab frame)~\cite{kontorovich1984}.
%the deviation in electronic energy due to phonons in the lab frame is given by
In Eq.~(\ref{eq:de}), ${\bf u}$ is the displacement vector for the ion at position ${\bf r}$ and time $t$, $\lambda_{i j}$ is the $(i,j)$ element of the deformation potential tensor ($i,j\in\{x,y,z\}$) and
\begin{equation}
  \tilde{\bf v}=\partial_{\bf p}\varepsilon_{\bf p}^{(0)} = \partial_{\bf p} \varepsilon_0 - \partial_{\bf p} ({\bf m}_{\bf p}\cdot{\bf B}) \equiv {\bf v} - \partial_{\bf p}({\bf m}_{\bf p}\cdot{\bf B})
\end{equation}
is the electronic velocity in a magnetic field and in the absence of lattice vibrations.
In the absence of a magnetic field, $\tilde{\bf v}$ becomes equal to ${\bf v}$.
Note that ${\bf v}$ and $\tilde{\bf v}$ are functions of ${\bf p}$; for brevity, we omit the momentum subscript.
Also for brevity, summations over repeated indices will be implicit throughout the text.
%Therefore, the dispersion of electrons takes the form

In a minimal model of a Weyl semimetal (WSM) containing two nodes of opposite chirality $\alpha=\pm 1$, separated in energy by $2\Delta$ ($\Delta\neq 0$ for chiral WSM), we have
\begin{align}
  \label{eo}
  \varepsilon_0^{(\alpha)}({\bf p}) &= v_F^{(\alpha)} p + \alpha\Delta \\
  \label{v0}
  {\bf v}^{(\alpha)} &= v_F^{(\alpha)} \hat{\bf p}\\ 
  \label{Om}
 \boldsymbol{\Omega}_{\textbf{p}}^{(\alpha)} &= |C|\alpha\hbar^{2} \frac{\hat{\bf p}}{2p^{2}}\\
 \label{m}
\textbf{m}_{\textbf{p}}^{(\alpha)} &= -|C|\alpha \hbar \frac{ev_{F}^{(\alpha)}}{2p}\hat{\bf p}\\
  \label{gradm}
  \partial_{\textbf{p}}(\textbf{m}^{(\alpha)}_{\bf p}\cdot\textbf{B}) &= -\alpha|C|\hbar\frac{ev^{(\alpha)}_{F}}{2p^{2}}\bigg(\textbf{B} - 2(\textbf{B}\cdot\hat{\bf p})\hat{\bf p}\bigg),
\end{align}
where the superscript $\alpha$ indicates that the momentum ${\bf p}$ is restricted to the vicinity of Weyl node $\alpha$ (not to be confused with the superscript $0$ appearing elsewhere), $v_F^{(\alpha)}$ is the Fermi velocity describing the slope of the energy dispersion at node $\alpha$, and $|C|=1$ is the Chern number.
Although $|C|=1$ for the minimal model, we will keep $|C|$ as a bookkeeping parameter for effects of geometric (Berry curvature, orbital magnetic moment) origin in the sound propagation.
In Eqs.~(\ref{eo}), (\ref{Om}), (\ref{m}) and (\ref{gradm}), ${\bf p}$ is the momentum measured with respect to the Weyl node.

For latter reference, the space and momentum derivatives of the energy dispersion are given as
\begin{align}
  \label{delp}
\partial_{\bf{p}} \varepsilon_{\textbf{p}}({\bf r},t)  &= \Tilde{\textbf{v}}+ u_{ij} \partial_{\textbf{p}}\left(\lambda_{ij} + p_{i} \Tilde{v}_{j}\right) + \dot{u}_{i} \partial_{\textbf{p}} p_{i} - m \dot{u}_{i} \partial_{\textbf{p}}\Tilde{v}_{i}\\
  \label{delr}
\partial_{\textbf{r}} \varepsilon_{\textbf{p}}({\bf r},t) &= \lambda_{ij}\partial_{\textbf{r}}u_{ij} + p_{i} \Tilde{v}_{j}\partial_{\textbf{r}}u_{ij} + p_{i}\partial_{\textbf{r}}{\dot{u_{i}}} - m\Tilde{v}_{i}\partial_{\textbf{r}}\dot{u_{i}}.
\end{align}
%with 
%\begin{align}
%    &\Tilde{\textbf{v}} = \partial_{\textbf{p}}\varepsilon_{p}^{(0)} = \textbf{v} - \partial_{\textbf{p}}(\textbf{m}\cdot\textbf{B})\\
%    &\textbf{v} = \partial_{\textbf{p}}\varepsilon_0.
%\end{align}

Plugging Eqs.~(\ref{delp}) and (\ref{delr}) in Eqs.~(\ref{r}) and (\ref{p}), we get
%the semiclassical equations of electronic motion in presence of lattice vibrations
\begin{align}
  \label{fp}
    \dot{\textbf{p}} &= \frac{e\textbf{E} + e \Tilde{\textbf{v}}\times\textbf{B} + \frac{e^2}{\hbar}\boldsymbol{\Omega}_{\textbf{p}}(\textbf{B}\cdot\textbf{E}) - \partial_{\textbf{r}}\delta\varepsilon+ e \partial_{\textbf{p}}\delta\varepsilon\times\textbf{B} - \frac{e}{\hbar}\boldsymbol{\Omega}_{\textbf{p}}(\textbf{B}\cdot\partial_{\textbf{r}}\delta\varepsilon)}{1+\frac{e}{\hbar}\textbf{B}\cdot\boldsymbol{\Omega}_{\textbf{p}}}\\
  \label{fr}
    \dot{\textbf{r}} &= \frac{\Tilde{\textbf{v}}+ \frac{e}{\hbar} {\bf E}\times {\bf\Omega}_{\bf p} + \frac{e}{\hbar}{\bf B} ({\bf \Omega}_{\bf p}\cdot  \Tilde{\textbf{v}}) + {\partial_{\textbf{p}}\delta\varepsilon} -\frac{1}{\hbar}\partial_{\textbf{r}}\delta\varepsilon\times {\bf \Omega}_{\bf p}+ \frac{e}{\hbar}{\bf B} ({\bf \Omega}_{\bf p}\cdot  {\bf \partial_{\textbf{p}}\delta\varepsilon})  }{1+\frac{e}{\hbar}\textbf{B}\cdot\boldsymbol{\Omega}_{\textbf{p}}}.
\end{align}

Next, we will use the equations of this section to set up the Boltzmann kinetic equation.

\subsection*{II. Boltzmann kinetic equation (BKE)}\label{sec:BKE}
The Boltzmann kinetic equation for the electronic distribution function $f_{\bf p}({\bf r},t)$   has the following form,
\begin{equation}\label{BKE}
    \partial_t f + {\bf\dot{r}}\cdot\partial_{\bf r} f +{\bf\dot{p}}\cdot\partial_{\bf p} f = I_{\rm coll} \{f\},
\end{equation}
where $I_{\rm coll}$ is the collision term.
We seek a solution of the form 
\begin{equation}\label{f}
    f_{\bf p}({\bf r},t) = f_{\bf p}^{\rm l. e.}(\textbf{r},t) + \chi_{\bf p}(\textbf{r},t)\frac{\partial f_{0}(\varepsilon_{\textbf{p}}^{(0)})}{\partial \varepsilon_{\textbf{p}}^{(0)}},
\end{equation}
where 
\begin{equation}\label{locf}
  \begin{split}
    f_{\bf p}^{\rm l. e.}(\textbf{r},t) \equiv f_{0}(\varepsilon_{\textbf{p}}({\bf r},t) - \textbf{p}\cdot\dot{\textbf{u}} -\mu_{0} -\delta\mu({\bf r},t))
\end{split}
\end{equation}
is the local equilibrium distribution function and $f_{0}$ is the Fermi-Dirac distribution.
The local equilibrium has an effective chemical potential $\mu_0+\delta\mu$, where $\mu_0$ is the chemical potential for $u=0$ and $\delta\mu$ is the correction due to lattice vibrations.
%evaluated in a magnetic field.
We will define $\delta\mu$ such that the following ``normalization condition'' is satisfied:
the electron density calculated from the local equilibrium distribution is equal to the total electron density~\cite{kontorovich1963, kontorovich1971}.
This condition implies a vanishing average of $\chi$ over the {\em total} Fermi surface of the system (which contains different pieces centered around different Weyl nodes) in the presence of a magnetic field.

%The normalization condition is not necessary to solve the BKE; one can proceed with the %calculation without adopting it and the final results should be independent from it. We %%instead regard the normalization condition as a tool that helps simplify certain %calculations.

The time derivative of the distribution function can be written as
\begin{equation}
\begin{aligned}\label{time}
    \partial_t f = \frac{\partial  f_{0}}{\partial  \varepsilon_{\bf p}^{(0)}}\left(\delta\dot{\varepsilon}- p_i\ddot{u}_i - \delta\dot{\mu} +\dot{\chi}\right) 
    = \frac{\partial  f_{0}}{\partial  \varepsilon_{\bf p}^{(0)}}\left(\lambda_{ij}\dot{u}_{ij} + p_{i}\Tilde{v}_{j}\dot{u}_{ij} - m\Tilde{v}_{i}{\ddot{u}}_{i} -{\delta}\dot{\mu}+\dot{\chi}\right). 
\end{aligned}
\end{equation}
Here and below, we omit the momentum subscript from $\chi$, for the sake of notational simplicity.
The space derivative is given as
\begin{equation}
\begin{aligned}\label{space}
  \partial_{\bf r} f& = \frac{\partial  f_{0}}{\partial  \varepsilon_{\bf p}^{(0)}}\left( \partial_{\textbf{r}}\delta\varepsilon - p_i\partial_{\textbf{r}}\dot{u}_i - \partial_{\textbf{r}}\delta\mu +\partial_{\bf r}\chi\right)
\end{aligned}
\end{equation}
and the momentum derivative has the form
\begin{equation}\label{momentum}
    \partial_{\bf p}f = \frac{\partial f_0}{\partial \varepsilon_{\bf p}^{(0)}} \bigg[\Tilde{\textbf{v}} + \partial_{\textbf{p}}\delta\varepsilon -\dot{u}_i \partial_{\bf p} p_i+\partial_{\bf p}\chi\bigg] + \Tilde{\textbf{v}}\bigg[ \delta\varepsilon - p_i \dot{u}_i-\delta\mu + \chi \bigg]\frac{\partial^2f_0}{\partial\varepsilon_{\bf p}^{(0)2}}.
\end{equation}
For the collision integral, we use the relaxation time approximation with two different relaxation times for the intravalley and intervalley scattering (see main text for an explanation):
\begin{equation}\label{coll}
I_{\rm coll}^{(\alpha)}\{f^{(\alpha)}\} =-\bigg[\Gamma_{A}\bigg(\chi^{(\alpha)} - \frac{\langle\chi^{(\alpha)}\rangle_{0}}{\langle1^{(\alpha)}\rangle}\bigg) +\Gamma_{E}\frac{\langle\chi^{(\alpha)}\rangle}{\langle1^{(\alpha)}\rangle}\bigg]\frac{\partial f_{0}^{(\alpha)}(\varepsilon_{\bf p}^{(0)})}{\partial \varepsilon_{\bf p}^{(0)}},
\end{equation} 
where $\alpha$ indicates that the momentum ${\bf p}$ is restricted to the vicinity of Weyl node $\alpha$.
As mentioned in the main text, the brackets $\langle ... \rangle$ denote an average over the equilibrium ($u=0$) Fermi surface (see also the next section of this supplemental material).

Plugging Eqs.~(\ref{fp}),~(\ref{fr}),~(\ref{time}),~(\ref{space}) and (\ref{momentum}) in Eq.~(\ref{BKE}), keeping terms up to first order in ${\bf u}$ and restricting ${\bf p}$ to the vicinity of node $\alpha$,  we get
\begin{align}\label{WSMBKE}
%\begin{split}
  &D^{(\alpha)}\partial_t\chi^{(\alpha)}+ \left(\Tilde{\textbf{v}}^{(\alpha)} + \frac{e}{\hbar}\textbf{B}(\boldsymbol{\Omega}^{(\alpha)}\cdot\Tilde{\textbf{v}}^{(\alpha)}) \right)\cdot\partial_{\bf r}\chi^{(\alpha)} + e(\Tilde{\textbf{v}}^{(\alpha)}\times\textbf{B})\cdot\partial_{\bf p}\chi^{(\alpha)}+ D^{(\alpha)}\bigg[\Gamma_{A}\left(\chi^{(\alpha)}-\frac{\langle\chi^{(\alpha)}\rangle}{\langle1^{(\alpha)}\rangle}\right)+ \Gamma_{E}\frac{\langle\chi^{(\alpha)}\rangle}{\langle1^{(\alpha)}\rangle}\bigg] \nonumber\\
  &= -D^{(\alpha)}\bigg(\lambda_{ij}^{(\alpha)}\dot{u}_{ij} + p_i\Tilde{v}_j^{(\alpha)}\dot{u}_{ij} - m\Tilde{v}_i^{(\alpha)}\ddot{u}_i - \delta\dot{\mu}\bigg) + \left(\Tilde{\textbf{v}}^{(\alpha)}+ \frac{e}{\hbar}(\boldsymbol{\Omega}^{(\alpha)}\cdot\Tilde{\textbf{v}}^{\alpha)})\textbf{B}\right)\cdot(p_i\partial_{\textbf{r}}\dot{u}_i + \partial_{\textbf{r}}\delta\mu) \nonumber\\
     & - \Tilde{\textbf{v}}^{(\alpha)}\cdot\left(e\textbf{E} + \frac{e^2}{\hbar}\boldsymbol{\Omega}^{(\alpha)}(\textbf{B}\cdot\textbf{E}) \right) + e(\Tilde{\textbf{v}}^{(\alpha)}\times\textbf{B})\cdot\dot{\textbf{u}},
%\end{split}
\end{align}
where 
\begin{equation}\label{D}
    D^{(\alpha)}\equiv 1+ \frac{e}{\hbar} {\bf{B\cdot\Omega^{(\alpha)}_{\textbf{p}}}}.
\end{equation}
Once again, for brevity we omit the momentum subscripts from quantities such as $\W$ and $D$.

Using $\partial_t\rightarrow -i\omega$, $\partial_{\bf r}\rightarrow i\textbf{q}$ and $\chi_{\bf p}(\textbf{r},t)\rightarrow \chi_{\bf p}(\textbf{q},\omega)$, the Fourier transform of Eq.~(\ref{WSMBKE}) reads
\begin{align}\label{WSMBKEf}
    &-i\omega D^{(\alpha)}\chi^{(\alpha)} + i\left( {\textbf{q}\cdot\Tilde{\textbf{v}}^{(\alpha)}} + \frac{e}{\hbar}(\textbf{q}\cdot\textbf{B})(\boldsymbol{\Omega}^{(\alpha)}\cdot\Tilde{\textbf{v}}^{(\alpha)})\right)\chi^{(\alpha)} + e(\Tilde{\textbf{v}}^{(\alpha)}\times\textbf{B})\cdot\partial_{\bf p}\chi^{(\alpha)}+ D^{(\alpha)}\bigg[\Gamma_{A}\chi^{(\alpha)}\nonumber\\
     & -\Gamma_{A}\frac{\langle\chi^{(\alpha)}\rangle}{\langle1^{(\alpha)}\rangle} + \Gamma_{E}\frac{\langle\chi^{(\alpha)}\rangle}{\langle1^{(\alpha)}\rangle}\bigg] = -D^{(\alpha)}\bigg[-i\omega\left(\lambda_{i j}^{(\alpha)} + p_i \tilde{v}_j^{(\alpha)}\right) u_{i j}
      + m\Tilde{v}^{(\alpha)}_i\omega^{2}u_{i} +i\omega\delta\mu\bigg]\nonumber\\ 
    &+ i (\delta\mu-i\omega p_{i}u_{i}) \left( {\textbf{q}\cdot\Tilde{\textbf{v}}^{(\alpha)}} + \frac{e}{\hbar}(\textbf{q}\cdot\textbf{B})(\boldsymbol{\Omega}^{(\alpha)}\cdot\Tilde{\textbf{v}}^{(\alpha)})\right) - \Tilde{\textbf{v}}^{(\alpha)}\cdot\left(e\textbf{E} + \frac{e^2}{\hbar}\boldsymbol{\Omega}^{(\alpha)}(\textbf{B}\cdot\textbf{E}) \right)
    - i\omega e(\Tilde{\textbf{v}}^{(\alpha)}\times\textbf{B})\cdot{\textbf{u}},
\end{align}
where the complex number $i$ should not be confused with the subscript $i$ appearing in $\lambda_{i j}$, $u_{i j}$, $u_i$ and $p_i$.
Also, $u_{i j} = i (q_i u_j + q_j u_i)/2$ is the Fourier transform of the strain tensor.
The momentum appearing in terms such as $p_i \tilde{v_j}$ and $p_i u_i$ can be decomposed as ${\bf p} = {\bf P}^{(\alpha)} + \delta{\bf p}$, where ${\bf P}^{(\alpha)}$ is the position of the Weyl node $\alpha$ in momentum space and $\delta{\bf p}$ is the momentum measured with respect to the node.
In the case of time-reversal-symmetric WSM that we will concentrate on hereafter, the contribution from terms involving ${\bf P}^{(\alpha)}$ will be cancelled between time-reversed partners.
Hence, in $p_i \tilde{v_j}$ and $p_i u_i$,  we can effectively think of ${\bf p}$ as the momentum measured with respect to Weyl node $\alpha$.

In the next section, we solve Eq.~(\ref{WSMBKEf}) perturbatively in the magnetic field.
First, we will solve the BKE for $B=0$ and then we will use this solution to derive an expression of $\chi$ valid to linear order of magnetic field. 

\subsection*{III. Solution of the BKE to zeroth order in magnetic field}
\label{sec:chi0}

The solution of Eq.~(\ref{WSMBKEf}) at $B=0$ will be labeled as $\chi_{0}$.
The electric field  and the shift of the chemical potential at $B=0$ will be likewise labelled as ${\bf E}_0$ and $\delta\mu_0$, respectively.
%The procedure includes finding expressions for the electric field and $\delta\mu$, which are unknown.
%Let us first write the expression for BKE in case of $\textbf{B}=\textbf{0}$.
%In this case, the electric field and $\delta\mu$ are also zeroth order in $B$, such that we will label them as $E_{0}$ and $\delta\mu_{0}$.
Also, we have $D^{(\alpha)} = 1$ and $\Tilde{\textbf{v}}^{(\alpha)}= \textbf{v}^{(\alpha)}$.
Therefore, Eq.~(\ref{WSMBKEf}) can be written as
\begin{equation}\label{WSMBKEB0}
\begin{split}
    (-i\omega + i\textbf{q}\cdot\textbf{v}^{(\alpha)}+\Gamma_{A})\chi_{0}^{(\alpha)} - (\Gamma_{A}-\Gamma_E)\frac{\langle\chi_{0}^{(\alpha)}\rangle_0}{\langle1^{(\alpha)}\rangle_0} 
    &=  -\lambda_{ij}^{(\alpha)}\omega q_{j}u_{i}-m v_i^{(\alpha)} \omega^2 u_i - i\omega\delta\mu_{0} - e{\bf E}_0\cdot{\bf v}^{(\alpha)}+ i\textbf{q}\cdot{\textbf{v}^{(\alpha)}}\delta\mu_{0},\\
\end{split}
\end{equation}
where the subscript $0$ in $\langle ... \rangle_0$ indicates that the average is taken on the $B=0$ Fermi surface.

Equation~(\ref{WSMBKEB0}) contains multiple unknowns: $\chi_0$, ${\bf E}_0$ and $\delta\mu_0$.
We will see that they can be related to one another by virtue of Maxwell's equations and the normalization condition.
To do so, we begin by recognizing that the total charge density can be written as $Q+ e n$, where
%we provide with a derivation of expression of $\delta\mu_{0}$. 
%Ionic charge density $Q$ (under deformation) is given as
\begin{equation}\label{ion}
    Q = -en_{0}(1-\partial_{\bf r}\cdot{\bf{u}})
\end{equation}
is the ionic charge (to first order in ${\bf u}$), $n_{0} = \langle\langle f_{0}\rangle\rangle_0$ is the electron density in the absence of lattice vibrations
and
\begin{equation}\label{n}
n =\langle\langle f \rangle \rangle_0
\end{equation}
is the total electron density.
Here, the double brackets $\langle\langle ... \rangle\rangle$ denote a momentum integral over the equilibrium ($u=0$) Brillouin zone, and the subscript $0$ in $\langle\langle ... \rangle\rangle_0$ is to remind that the integral is done for $B=0$.

Therefore, the total charge density is 
\begin{equation}\label{charge}
\begin{split}
    Q+en &= -en_{0}(1-\partial_{\bf r}\cdot{\bf{u}}) + e\langle\langle f\rangle \rangle_0\\
    & = -en_{0} + en_{0} \partial_{\bf r} \cdot{\bf{u}}+ e\langle\langle f_{0}\rangle\rangle_0 -e\langle\lambda_{i j} u_{i j} + p_i v_j u_{i j} - m v_i \dot{u}_i -\delta\mu_0 +\chi_0\rangle_0\\
    &= en_{0} \partial_{\bf r}\cdot{\bf{u}} -e[\langle\lambda_{ij}\rangle_0 u_{ij} + (\partial_{\bf r}\cdot {\bf{u}})n_{0} -\delta\mu_{0}\langle1\rangle_0]-e\langle\chi_{0}\rangle_0\\
    &= -e\langle\lambda_{ij}\rangle_0 u_{ij} + e\delta\mu_{0}\langle1\rangle_0-e\langle\chi_{0}\rangle_0,
\end{split}
\end{equation}
where we have used $\langle O_{\bf p} \rangle_0 =0$ for any function $O_{\bf p}$ that is odd in momentum, and $\langle p_i v_j\rangle_0 = \delta_{i j} n_0$.
We remind the reader that the notation $\langle ... \rangle$ implies an average over the full Fermi surface, while $\langle ...^{(\alpha)} \rangle$ indicates an average over the piece of the Fermi surface surrounding the node $\alpha$.
Evidently, $\langle...\rangle=\sum_\alpha\langle...^{(\alpha)}\rangle$.

According to Gauss's law, 
\begin{equation}\label{gauss}
\epsilon_{lat} \partial_{\bf r}\cdot {\bf E}_0 = Q+en,
\end{equation}
where $\epsilon_{lat}$ is the high-frequency dielectric permittivity.
Plugging Eq.~(\ref{gauss}) in Eq.~(\ref{charge}), we get
\begin{equation}\label{gauss1}
%\begin{split}
\epsilon_{lat} \partial_{\bf r}\cdot {\bf E}_{0} = -e\langle\lambda_{ij}\rangle_0 u_{ij} +  e\delta\mu_{0}\langle1\rangle_0-e\langle\chi_{0}\rangle_0.\\
%\end{split}
\end{equation}
At zero magnetic field, the normalization condition implies $\langle\chi_{0}\rangle_0 =0$.
Then, 
\begin{equation}\label{deltamu}
\delta{\mu}_{0} = \frac{\epsilon_{lat} \partial_{\bf r}\cdot \textbf{E}_{0}}{ e\langle1\rangle_0}  + \frac{\langle\lambda_{ij}\rangle_0}{\langle1\rangle_0}u_{ij}.\\
\end{equation}
Fourier transforming,
\begin{equation}\label{FTdelmunew}
\delta{\mu}_{0} = \frac{i\textbf{q}\cdot \textbf{E}_{0} \epsilon_{lat}}{ e\langle1\rangle_0}  + \frac{\langle\lambda_{ij}\rangle_0}{\langle1\rangle_0}iq_{j}u_{i}.\\
\end{equation}
Replacing Eq.~(\ref{FTdelmunew}) in Eq.~(\ref{WSMBKEB0}) yields
\begin{equation}\label{chi0B0E0new}
\begin{split}
  \chi_{0}^{(\alpha)} = R^{(\alpha)} \bigg[ &-\lambda_{ij}^{(\alpha)}\omega q_{j}u_{i}+ \frac{\langle\lambda_{ij}\rangle_0}{\langle 1\rangle_0}\omega q_{j}u_{i}-m \omega^2 v_i^{(\alpha)} u_i - ({\textbf{q}}\cdot{\textbf{v}^{(\alpha)}})\frac{\langle\lambda_{ij}\rangle_0}{\langle 1\rangle_0}q_{j}u_{i} - e\textbf{E}_{0}\cdot{\textbf{v}^{(\alpha)}} + \frac{\omega (\textbf{q}\cdot\textbf{E}_{0})\epsilon_{lat}}{e\langle1\rangle_0}\\
  &-\frac{(\textbf{q}\cdot{\textbf{v}^{(\alpha)})(\textbf{q}\cdot\textbf{E}_{0})}\epsilon_{lat}}{e\langle1\rangle_0} 
+ (\Gamma_{A}-\Gamma_E)\frac{\langle\chi_{0}^{(\alpha)}\rangle_0}{\langle1^{(\alpha)}\rangle_0} \bigg],\\
\end{split}
\end{equation}
where
\begin{equation}
  R^{(\alpha)} = (-i\omega + i{\textbf{q}}\cdot{\textbf{v}^{(\alpha)}} +\Gamma_{A})^{-1}.
\end{equation}

We have thus removed one of the unknowns ($\delta\mu_0$).
We still need to determine the electric field.
To do so, we begin by taking the average of Eq.~(\ref{chi0B0E0new}) over the Fermi surface surrounding the node $\alpha$.
The calculation becomes simple if the following conditions are simultaneously satisfied:
\begin{equation}
  \label{eq:conds}
  {\rm (i)}\, \Gamma_A\gg v_F^{(\alpha)} q,\,\, {\rm (ii)}\, \Gamma_E\gg {\rm max}\left(\omega, \frac{(v_F^{(\alpha)})^2 q^2}{\Gamma_A}\right).
\end{equation}
These conditions are realistic for long-wavelength acoustic phonons.
For instance, for a phonon wave vector $q<5\times10^5\, {\rm m^{-1}}$, we have $\hbar v q\lesssim 10^{-2}\, {\rm meV}$ and $\hbar \omega \lesssim 10^{-4}\, {\rm meV}$.
Both of these energy scales are small compared to $\hbar\Gamma_A$ in reasonably disordered WSM (where one may anticipate $\hbar\Gamma_A \sim 1-10\, {\rm meV}$ at low temperatures).
In contrast, $\hbar\Gamma_E$ has been estimated to be of the order of $10^{-2}\, {\rm meV}$ because intervalley impurity scattering is suppressed with respect to intravalley impurity scattering.
The fact that $\Gamma_A\gg\Gamma_E$ will be exploited in the next sections for further simplifications.

If the conditions (i) and (ii) above are realized, it is safe to approximate 
%We apply another energy scaling with respect to R. Since $\hbar qv \approx 10^{-3}$ meV, $\hbar\omega = 10^{-5}$ meV, therefore,
\begin{equation}\label{Rapprox}
R^{(\alpha)} \approx \frac{1}{\Gamma_A}.
\end{equation}
With this proviso, Eq.~(\ref{chi0B0E0new}) becomes
\begin{equation}\label{E0chi0}
\begin{split}
  \chi_{0}^{(\alpha)}&= -\frac{\lambda_{ij}^{(\alpha)}\omega q_{j}u_{i}}{\Gamma_{A}}+ \frac{\langle\lambda_{ij}\rangle_0}{\langle 1\rangle_0}\frac{\omega q_{j}u_{i}}{\Gamma_{A}}-\frac{m \omega^2  v_i^{(\alpha)} u_i}{\Gamma_A} +\frac{\omega\epsilon_{lat}(\textbf{q}\cdot\textbf{E}_{0})}{e\langle1\rangle_0\Gamma_{A}} -\frac{e\textbf{E}_{0}\cdot\textbf{v}}{\Gamma_{A}} - \frac{\textbf{q}\cdot\textbf{v}}{\Gamma_{A}}\frac{\langle\lambda_{ij}\rangle_0}{\langle1\rangle_0}q_{j}u_{i}-\frac{(\textbf{q}\cdot\textbf{v})(\textbf{q}\cdot\textbf{E}_{0})\epsilon_{lat}}{e\langle1\rangle_0\Gamma_{A}}\\
&\quad + \bigg(1 - \frac{\Gamma_{E}}{\Gamma_{A}}\bigg)\frac{\langle\chi_{0}^{(\alpha)}\rangle_0}{\langle1^{(\alpha)}\rangle_0}.
\end{split}
\end{equation}
In the limit $\Gamma_A\to\infty$, $\chi_0\to \langle \chi_0^{(\alpha)}\rangle/\langle 1^{(\alpha)}\rangle$ becomes independent of the direction of momentum.
This limit could have been anticipated from the form of the collision term.
Below, the limit $\Gamma_A\to\infty$ will be useful to extract the leading terms contributing to the phonon magnetochiral effect.

Now we take Fermi surface averages of Eq. (\ref{E0chi0})  around each node. It is here that the condition $\Gamma_E\gg {\rm max}(\omega, (v_F^{(\alpha)})^2 q^2/\Gamma_A)$ is invoked.
Indeed, without any approximations, one has
\begin{equation}
  \langle R^{(\alpha)}\rangle_0 =\frac{\langle 1^{(\alpha)}\rangle_0}{2 q v_F^{(\alpha)}}\left[\tan^{-1}\left(\frac{v_F^{(\alpha)} q-\omega}{\Gamma_A}\right) + \tan^{-1}\left(\frac{v_F^{(\alpha)} q+\omega}{\Gamma_A}\right)\right] + i \frac{\langle 1^{(\alpha)}\rangle_0}{4 q v_F^{(\alpha)}} \ln\left[\frac{(q v_F^{(\alpha)} +\omega)^2+\Gamma_A^2}{(q v_F^{(\alpha)}-\omega)^2+\Gamma_A^2}\right],
\end{equation}
which, for $\Gamma_A\gg v_F^{(\alpha)} q \gg \omega$, gives
\begin{equation}
  \langle R^{(\alpha)}\rangle_0\simeq \frac{\langle 1^{(\alpha)}\rangle_0}{\Gamma_A}\left(1-\frac{(v_F^{(\alpha)})^2 q^2}{3 \Gamma_A^2}+\frac{i\omega}{\Gamma_A}\right).
\end{equation}
The Fermi surface average of the last term in the right hand side of Eq.~(\ref{chi0B0E0new}) then gives
\begin{equation}
  \label{eq:ela}
  \langle\chi_0^{(\alpha)}\rangle_0\left(1-\frac{\Gamma_E}{\Gamma_A}\right)\left(1-\frac{(v_F^{(\alpha)})^2 q^2}{3 \Gamma_A^2}+\frac{i\omega}{\Gamma_A}\right)\simeq \langle\chi_0^{(\alpha)}\rangle_0\left(1-\frac{\Gamma_E}{\Gamma_A}-\frac{(v_F^{(\alpha)})^2 q^2}{3 \Gamma_A^2}+\frac{i\omega}{\Gamma_A}\right).
\end{equation}
The first term in Eq.~(\ref{eq:ela}) is clearly the largest; however, it cancels with the average of the left hand side of Eq.~(\ref{chi0B0E0new}).
Doing the approximation in Eq.~(\ref{Rapprox}) is tantamount to saying that, in Eq.~(\ref{eq:ela}), $\Gamma_E/\Gamma_A \gg (v_F^{(\alpha)})^2 q^2/\Gamma_A^2$ and $\Gamma_E/\Gamma_A \gg i\omega/\Gamma_A$, so that the last two terms in Eq. (\ref{eq:ela}) may be neglected. This then results in the conditions shown in Eq.~(\ref{eq:conds}).

Taking the Fermi surface averages of Eq. (\ref{E0chi0}) around each node results in a system of two equations which, combined with the normalization condition $\langle\chi_{0}^{(+)}\rangle_0 + \langle\chi_{0}^{(-)}\rangle_0 =0$, gives
\begin{equation}\label{E0}
E_{0,\parallel} = 0
\end{equation}
for the longitudinal component of the electric field, and
\begin{equation}\label{avchi0}
  \langle\chi_{0}^{(+)}\rangle_0 = -\langle\chi_{0}^{(-)}\rangle_0=-\frac{\omega q_{j}u_{i}}{\Gamma_{E}\langle1\rangle_0}\bigg[\langle\lambda_{ij}^{(+)}\rangle_0\langle1^{(-)}\rangle_0 -\langle\lambda_{ij}^{(-)}\rangle_0\langle1^{(+)}\rangle_0\bigg].
\end{equation}
We note that the longitudinal part of the electric field would not have vanished if we had not approximated $R^{(\alpha)}$ as $1/\Gamma_A$.
In the isotropic approximation for the deformation potential tensor ($\lambda_{i j} = \lambda_1 \delta_{i j} + \lambda_2 p_i p_j/p^2$), we find
\begin{equation}
  \langle\chi_{0}^{(+)}\rangle_0 = -\langle\chi_{0}^{(-)}\rangle_0=-\frac{\omega q_{i}u_{i}\langle 1^{(+)}\rangle_0\langle 1^{(-)}\rangle_0}{\Gamma_{E}\langle1\rangle_0}\left(\lambda_1^{(+)}-\lambda_1^{(-)}+\frac{\lambda_2^{(+)}}{3}-\frac{\lambda_2^{(-)}}{3}\right),
\end{equation}
where we have used $\langle \lambda_{i j}^{(\alpha)}\rangle_0 = \delta_{i j} (\lambda_1^{(\alpha)} + \lambda_2^{(\alpha)}/3) \langle 1^{(\alpha)}\rangle$.
We thus learn that, in the absence of a magnetic field, it is necessary to have a different deformation potentials on Weyl nodes of opposite chirality in order for lattice vibrations to induce an electrochemical potential difference between them.
This occurs only in chiral WSM, where $\lambda_{1(2)}^{(+)} \neq \lambda_{1 (2)}^{(-)}$ and nonchiral crystals with pseudoscalar acoustic phonons (see main text).
%It is implicit that when we write $\langle\rangle^{(+)}$, all terms inside are around $(+)$ node and similarly $\langle\rangle^{(+)}$ implies, all terms inside are around $(-)$ node.

The transverse component of the electric field may be obtained by combining Faraday's law with Amp\`ere-Maxwell's law, and by computing the current density from the electronic distribution function.
For simplicity, we neglect the transverse electric fields produced by sound waves.
This may be justified by the fact that the magnetic fields induced by lattice vibrations are small and vary slowly in time.

Therefore, Eq.~(\ref{E0chi0}) becomes
%we write an expression of $\chi_{0}$ as
\begin{equation}\label{chi0finale}
\chi_{0}^{(\alpha)}= -\frac{\omega}{\Gamma_{A}}q_{j}u_{i}\bigg(\lambda_{ij}^{(\alpha)}-\frac{\langle\lambda_{ij}\rangle_0}{\langle1\rangle_0}\bigg) - \frac{ m \omega^2 v_i^{(\alpha)} u_i}{\Gamma_A}-\frac{\textbf{q}\cdot\textbf{v}}{\Gamma_{A}}\frac{\langle\lambda_{ij}\rangle_0}{\langle1\rangle_0} q_{j} u_{i} + \frac{\langle\chi_{0}^{(\alpha)}\rangle_0}{\langle1^{(\alpha)}\rangle_0}\bigg(1-\frac{\Gamma_{E}}{\Gamma_{A}}\bigg).
\end{equation}
This completes the approximate solution of the BKE to zeroth order in $B$.
It contains no signatures of the Berry curvature.
Such signatures will only appear when we turn on the magnetic field.
Next, we will search for the solution of the BKE to first order in $B$.
%In our next section, we will show the derivation of $\chi_{1}$ which is the solution of BKE in presence of magnetic field. 

\subsection*{IV. Solution of the BKE to first order in the magnetic field}
\label{sec:chi1}

%In this section, we will first derive a definition of Fermi surface averages in presence of magnetic field. This definition is crucial to the solution of BKE in presence of magnetic field.
In this section, we derive the solution of the BKE to first order in a magnetic field. 
Before embarking on the subject, we present some mathematical preliminaries that will prove useful later on.
We begin by recalling that, in a magnetic field, the expression for density of states in momentum space changes as 
\begin{equation}
    \frac{1}{(2\pi\hbar)^3}\longrightarrow \frac{1}{(2\pi\hbar)^3}(1 + \frac{e}{\hbar}\textbf{B}\cdot\boldsymbol{\Omega}_{\textbf{p}}).
\end{equation}
Accordingly, the volume integral of a function $\psi_{\bf p}({\bf B})$ over the Brillouin zone reads
\begin{equation}
  \langle\!\langle \psi_{\bf p}({\bf B})\rangle\!\rangle = \int \frac{d^3 p}{(2\pi\hbar)^3} \psi_{\bf p}({\bf B}) \left(1+\frac{e}{\hbar} {\bf B}\cdot\W_{\bf p}\right).
\end{equation}
Similarly, the Fermi surface average of $\psi_{\bf p} ({\bf B})$ reads
\begin{equation}\label{vol}
\langle \psi_{\bf p}({\bf B})\rangle = -\langle\!\langle\psi_{\textbf{p}}(\textbf{B})\frac{\partial f_0}{\partial\varepsilon_{\textbf{p}}^{(0)}}\rangle\!\rangle = \int\frac{d^{3}p}{(2\pi\hbar)^{3}}\psi_{\textbf{p}}(\textbf{B})\left(1+\frac{e}{\hbar} {\bf{B}\cdot\Omega_{\textbf{p}}}\right)\delta(\varepsilon_{\bf p}^{(0)}-\mu_{0}).
\end{equation}
Below, we will be interested in evaluating volume and surface integrals to first order in magnetic field.
We denote these quantities as $\langle\!\langle \psi_{\bf p}({\bf B})\rangle\!\rangle_1$ and $\langle \psi_{\bf p}({\bf B})\rangle_1$, respectively.
The formal expression for $\langle\!\langle \psi_{\bf p}({\bf B})\rangle\!\rangle_1$ can be rapidly obtained:
\begin{align}
  \label{briB}
  \langle\!\langle \psi_{\bf p}({\bf B})\rangle\!\rangle_1 &\simeq \int \frac{d^3 p}{(2\pi\hbar)^3} \left(\psi_{\bf p}(B=0)+{\bf B}\cdot\partial_{\bf B} \psi_{\bf p}({\bf B})|_{B=0}\right) \left(1+\frac{e}{\hbar} {\bf B}\cdot\W_{\bf p}\right)\nonumber\\
  &\simeq \langle\!\langle \psi_{\bf p}(B=0)\rangle\!\rangle_0 +
  \langle\!\langle {\bf B}\cdot\partial_{\bf B} \psi_{\bf p}({\bf B})|_{B=0}\rangle\!\rangle_0
  + \frac{e}{\hbar} \langle\!\langle \psi_{\bf p}(B=0) \W_{\bf p}\cdot{\bf B}\rangle\!\rangle_0.
\end{align}
The formal expression for $\langle\psi_{\bf p}({\bf B})\rangle_1$ is slightly more cumbersome due to the presence of the Dirac delta and the fact that the energy of the electrons depends on the magnetic field via the magnetic moment ${\bf m}_{\bf p}$:
\begin{equation}\label{setavB1}
\begin{aligned}
  \langle\psi^{(\alpha)}_{\textbf{p}}(\textbf{B})\rangle_{1} \simeq \int\frac{d^3p}{(2\pi\hbar)^3}\left(\psi^{(\alpha)}_{\textbf{p}}({\textbf{B}}=0) + \textbf{B}\cdot\partial_{\bf B}\psi^{(\alpha)}_{\textbf{p}}({\bf B})|_{B=0}
  \right)\left(1 + \frac{e}{\hbar}\boldsymbol{\Omega}^{(\alpha)}_{\textbf{p}}\cdot\textbf{B}\right)\times\\ \times\left( \delta(\varepsilon^{(\alpha)}_0({\bf p}) - \mu_0) - \frac{\textbf{m}^{(\alpha)}_{\textbf{p}}\cdot\textbf{B}}{v_F^{(\alpha)}}\partial_p\delta(\varepsilon^{(\alpha)}_0({\bf p})-\mu_{0})\right),
\end{aligned}
\end{equation}
where, as usual, the subscript $\alpha$ is to remind that the momentum integral in Eq. (\ref{setavB1}) is restricted to the vicinity of the Weyl node $\alpha$.
In the last line of Eq. (\ref{setavB1}), we have used
\begin{equation}
  \frac{\partial F(\varepsilon^{(\alpha)}_0)}{\partial \varepsilon^{(\alpha)}_0} = \left(\frac{\partial\varepsilon^{(\alpha)}_0}{\partial p}\right)^{-1} \frac{\partial F (\varepsilon_0^{(\alpha)})}{\partial p}= \frac{1}{v_F^{(\alpha)}} \frac{\partial F (\varepsilon^{(\alpha)}_0)}{\partial p},
\end{equation}
valid for any function $F$ that depends on momentum only via $\varepsilon^{(\alpha)}_0 = v_F^{(\alpha)} p$.
Here, $v_F^{(\alpha)}$ is the Fermi velocity or the slope of the Weyl dispersion in the vicinity of node $\alpha$, and $p$ is the magnitude of the momentum measured from the node.
Neglecting $O(B^2)$ terms in Eq.~(\ref{setavB1}), we have
\begin{equation}\label{averageB1}
\begin{aligned}
    \langle\psi^{(\alpha)}_{\textbf{p}}(\textbf{B})\rangle_{1} &= \int\frac{d^3p}{(2\pi\hbar)^3 v_F^{(\alpha)}}\left[\psi^{(\alpha)}_{\textbf{p}}({\bf{B}}=0)+\textbf{B}\cdot\partial_{\bf B}\psi^{(\alpha)}_{\textbf{p}}({\bf B})|_{B=0} +  \frac{e}{\hbar}\psi^{(\alpha)}_{\textbf{p}}({\bf{B}} =0)(\boldsymbol{\Omega}^{(\alpha)}_{\textbf{p}}\cdot\textbf{B}) \right]\delta(p-p_F^{(\alpha)})\\
    \\&-\int\frac{d^3p}{(2\pi\hbar)^3}\psi^{(\alpha)}_{\textbf{p}}{(\bf{B}} =0) \frac{\textbf{m}^{(\alpha)}_{\textbf{p}}\cdot\textbf{B}}{(v_F^{(\alpha)})^2}\partial_p\delta(p-p^{(\alpha)}_F),
\end{aligned}
\end{equation}
where $p_F^{(\alpha)}$ is the Fermi momentum measured from node $\alpha$.
From Eq.~(\ref{eo}), $p_F^{(\alpha)}$ is defined via $\mu_0=v_F^{(\alpha)} p_F^{(\alpha)} + \alpha\Delta$.
Using the identity
\begin{equation}\label{deltadef}
\int f(p) \partial_p \delta (p-p_0) dp = - \partial_p f(p)|_{p=p_0},
\end{equation}
we rewrite Eq.~(\ref{averageB1}) as
\begin{equation}\label{averagefinalB1_0}
\begin{split}
  \langle\psi^{(\alpha)}_{\textbf{p}}(\textbf{B})\rangle_{1} &= \int\frac{dS^{(\alpha)}_F}{(2\pi\hbar)^3 v_F^{(\alpha)}}\left[\psi^{(\alpha)}_{\textbf{p}_F}({\bf{B}} =0) + \textbf{B}\cdot\partial_{\bf B}\psi^{(\alpha)}_{\textbf{p}_F}({\bf B})\Big|_{B=0} + \frac{e}{\hbar}\psi_{\textbf{p}_F}^{(\alpha)}({\textbf{B}} = 0)\boldsymbol({\Omega}^{(\alpha)}_{\textbf{p}_F}\cdot\textbf{B})\right]\\
  &+\int \frac{\sin\theta d\theta d\phi}{(2\pi\hbar)^{3} (v_{F}^{(\alpha)})^{2}}\partial_p \left(p^2\psi_{\textbf{p}}{(\textbf{B}} = 0)\textbf{m}_{\textbf{p}}\cdot{\bf B}\right)\Big|_{p=p_F^{(\alpha)}},
\end{split}
\end{equation}
where $(\theta,\phi)$ are the polar and azimuthal angles in spherical coordinates and
\begin{equation}\label{FS}
    dS^{(\alpha)}_{F} = (p^{(\alpha)}_{F})^{2} \sin\theta d\theta d\phi
\end{equation}
is the surface area element on the Fermi surface near node $\alpha$. 
In other words,
\begin{equation}
  \label{averagefinalB1}
  \begin{split}
    \langle\psi^{(\alpha)}_{\textbf{p}}(\textbf{B})\rangle_{1}&=\langle\psi^{(\alpha)}_{\textbf{p}}({\bf{B}} = 0)\rangle_0 + \bigg\langle \frac{e}{\hbar}(\textbf{B}\cdot\boldsymbol{\Omega}^{(\alpha)}_{\textbf{p}}\psi^{(\alpha)}_{\textbf{p}}({\bf{B}} =0)\bigg\rangle_0+ \bigg\langle {\bf B}\cdot\left(\partial_{\bf B} \psi^{(\alpha)}_{{\bf p}}\right)\Big|_{B=0}\bigg\rangle_0 \\
  &+ \frac{1}{v_F^{(\alpha)} (p_F^{(\alpha)})^2}\bigg\langle\partial_p \left(p^2\psi^{(\alpha)}_{\textbf{p}}{(\textbf{B}} = 0)\textbf{m}^{(\alpha)}_{\textbf{p}}\cdot{\bf B}\right)\bigg\rangle_0,
\end{split}
\end{equation}
where
\begin{equation}\label{notationaverage}
\langle \psi_{\bf p}^{(\alpha)}\rangle_0 = \frac{1}{(2\pi\hbar)^{3}}\int\frac{dS^{(\alpha)}_{F}}{v^{(\alpha)}_{F}} \psi_{{\bf p}_F^{(\alpha)}}^{(\alpha)}.
\end{equation}
In our minimal model of electron-doped WSM, for which  ${\bf m}_{\bf p}^{(\alpha)} = -e v_F^{(\alpha)} p \W_{\bf p}^{(\alpha)}$, Eq.~(\ref{averagefinalB1}) can be further simplified as
\begin{equation}
    \langle\psi^{(\alpha)}_{\textbf{p}}(\textbf{B})\rangle_{1}=\langle\psi^{(\alpha)}_{\textbf{p}}({\bf{B}} = 0)\rangle_0 +  \bigg\langle {\bf B}\cdot\left(\partial_{\bf B} \psi^{(\alpha)}_{{\bf p}}\right)\Big|_{B=0}\bigg\rangle_0 
  + \frac{1}{v_F^{(\alpha)}}\bigg\langle \left({\bf m}_{{\bf p}}^{(\alpha)}\cdot{\bf B}\right) \left(\partial_p \psi_{\bf p}^{(\alpha)}(B=0)\right)\bigg\rangle_0.
\end{equation}

Armed with Eqs. (\ref{briB}) and (\ref{averagefinalB1}), we now begin to derive the solution of the BKE in the presence of magnetic field.
To linear order in $B$, we can expand
\begin{equation}\label{pert}
\begin{split}
    &\chi= \chi_0 + \chi_1 \\
    &\delta\mu= \delta\mu_0 + \delta\mu_1\\
    &\textbf{E}= \textbf{E}_0 + \textbf{E}_1,
\end{split}
\end{equation}
where the subscripts $0$ and $1$ have the meanings of zeroth order and linear order in $B$, respectively.
Using Eq.~(\ref{pert}) and Eq.~(\ref{WSMBKEB0}), we collect terms that are first order in $B$ and arrive at

\begin{align}\label{simpliWSMBKEB1E0}
   & \frac{\chi_{1}^{(\alpha)}}{R^{(\alpha)}}= -\frac{e}{\hbar}(\textbf{B}\cdot\boldsymbol{\Omega}^{(\alpha)})\bigg[\lambda_{ij}^{(\alpha)}\omega q_{j}u_{i} -\frac{\langle\lambda_{ij}\rangle_0}{\langle1\rangle_0}\omega q_{j}u_{i} + \chi_{0}^{(\alpha)}(\Gamma_{A}-i\omega) - (\Gamma_{A}-\Gamma_E)\frac{\langle\chi_{0}^{(\alpha)}\rangle_0}{\langle1^{(\alpha)}\rangle_0}\bigg]\nonumber \\
  %+ \Gamma_{E}\frac{\langle\chi_{0}^{(\alpha)}\rangle_0}{\langle 1^{(\alpha)}\rangle_0}\bigg]
  &+ i\textbf{q}\cdot\bigg(\frac{e}{\hbar}(\boldsymbol{\Omega}^{(\alpha)}\cdot\textbf{v}^{(\alpha)})\textbf{B} -\partial_{\textbf{p}}(\textbf{m}^{(\alpha)}\cdot\textbf{B}) \bigg)\bigg(iq_{j}u_{i}\frac{\langle\lambda_{ij}\rangle_0}{\langle1\rangle_0} -\chi_{0}^{(\alpha)}\bigg)-e(\textbf{v}^{(\alpha)}\times\textbf{B})\cdot(i\omega \textbf{u}+\partial_{\bf p}\chi_0^{(\alpha)})\nonumber \\
  &+ i({\textbf{q}}\cdot{\textbf{v}^{(\alpha)}}-\omega)\delta\mu_{1}- e\textbf{v}^{(\alpha)}\cdot\textbf{E}_1 + (\Gamma_{A}-\Gamma_E)\frac{\langle\chi_{1}^{(\alpha)}\rangle_0}{\langle1^{(\alpha)}\rangle_0},
\end{align}
where $\chi_0$ may be replaced by Eq.~(\ref{chi0finale}).
As expected, $\chi_1^{(\alpha)}\to \langle\chi_1^{(\alpha)}\rangle_0/\langle 1^{(\alpha)}\rangle_0$ when $\Gamma_A\to\infty$.
In the derivation of Eq. (\ref{simpliWSMBKEB1E0}), we have used ${\bf E}_0\simeq 0$ and $\delta\mu_0\simeq i q_j u_i \langle\lambda_{i j}\rangle_0/\langle 1\rangle_0$.
We have also used the relations
\begin{align}
  \label{eq:rels}
  \langle \chi_0^{(\alpha)} + \chi_1^{(\alpha)}\rangle_1 &\simeq \langle \chi_0^{(\alpha)}\rangle_1+\langle\chi_1^{(\alpha)}\rangle_0 \simeq \langle \chi_0^{(\alpha)}\rangle_0+\langle \chi_1^{(\alpha)}\rangle_0\nonumber\\
  \langle \lambda_{i j}\rangle_1 &\simeq \langle \lambda_{i j}\rangle_0\nonumber\\
  \langle 1^{(\alpha)}\rangle_1 &\simeq \langle 1^{(\alpha)}\rangle_0,
\end{align}
which are valid to first order in $B$ and can be obtained from Eq.~(\ref{averagefinalB1}).
The last equality in the first line of Eq.~(\ref{eq:rels}) relies on the fact that $\partial_p \chi_0^{(\alpha)}$ is even in momentum (note that $\partial_p$ is the derivative with respect to the {\em magnitude} of the momentum).
In addition, in Eq.~(\ref{simpliWSMBKEB1E0}) we have omitted certain terms that are small and make a negligible contribution to the final results.
These omissions rely on the fact that the following dimensionless ratios are very small for weakly doped semimetals:
\begin{equation}
  \frac{p_F^{(\alpha)} v_F^{(\alpha)}}{\lambda^{(\alpha)}} , \,\, \frac{m v_F^{(\alpha)} \omega^2}{\lambda^{(\alpha)} \omega q}, \,\,\frac{c_s p_F^{(\alpha)}}{\lambda^{(\alpha)}}, \,\, \frac{m c_s^2}{\lambda^{(\alpha)}}.
\end{equation}
Here, $c_s$ is the speed of sound in the absence of itinerant electrons and $\omega\simeq c_s q$ (modulo small corrections that we aim to calculate below).

Equation~(\ref{simpliWSMBKEB1E0}) contains various unknowns: $\chi_1^{(\alpha)}$, $\delta\mu_1$ and ${\bf E}_1$.
They are related to one another by virtue of Maxwell's equations and the normalization condition.
The procedure to find these relations is akin to the one followed for the $B=0$ case.
%Much like in the previous section, we us
Like in that case, we will apply the approximation  $R^{(\alpha)}\simeq 1/\Gamma_{A}$, which is justified when $\Gamma_A \gg v_F^{(\alpha)} q$ and $\Gamma_E\gg {\rm max}(\omega,(v_F^{(\alpha)})^2 q^2/\Gamma_A)$.

%In presence of magnetic field, we follow the same formalism as before to derive an expression of $\delta\mu_{1}$ and $E_{1}$. Recognizing that ionic charge density does not vary with first order in magnetic field, we can write
To first order in $B$ and $u$, the total charge density reads
\begin{equation}\label{totalchargeB}
\begin{split}
Q + en &= -e\langle\!\langle f_0\rangle\!\rangle_1(1-\partial_{\bf r}\cdot \textbf{u}) + e\langle\!\langle f\rangle\!\rangle_{1}\\
& \quad = en_{0}\partial_{\bf r}\cdot \textbf{u} - e\langle\lambda_{ij}u_{ij}+p_i \tilde{v}_j u_{i j} - m \tilde{v}_i\dot{u}_i -\delta\mu_{0}-\delta\mu_{1}+\chi_0+\chi_1\rangle_1.
\end{split}
\end{equation}
The normalization condition implies that $\langle \chi_0+\chi_1\rangle_1=0$.
Also, to first order in $B$, $\langle\!\langle f_0\rangle\!\rangle_1\simeq \langle\!\langle f_0\rangle\!\rangle_0$, $\langle p_i \tilde{v}_j\rangle_1\simeq \langle p_i v_j\rangle_0=\delta_{i j}\langle\!\langle f_0\rangle\!\rangle_0$, $\langle\lambda_{i j}\rangle_1\simeq \langle \lambda_{ i j}\rangle_0$ and $\langle 1\rangle_1\simeq \langle 1\rangle_0$.
Using these relations, together with ${\bf E}_0\simeq 0$ and $\delta\mu_0\simeq u_{ i j}\langle\lambda_{i j}\rangle_0/\langle 1\rangle_0$, Gauss' law can be written as
\begin{equation}\label{delmuE1}
\epsilon_{lat} \partial_{\bf r}\cdot \textbf{E}_{1} = e \delta\mu_1 \langle 1\rangle_0+ e m \dot{u}_i \langle \tilde{v}_i\rangle_1,
\end{equation}
%Here, $\langle\lambda_{ij}\rangle_{1} = \langle\lambda_{ij}\rangle$ (using Eq.~(\ref{averagefinalB1})) and using Fourier transform, the above equation reduces to
which yields
\begin{equation}\label{delmu1finale}
\begin{split}
\delta\mu_{1} &= \frac{i\textbf{q}\cdot\textbf{E}_{1} \epsilon_{lat}}{e\langle1\rangle_0}+\frac{i m \omega \langle \tilde{v}_i\rangle_1 u_i}{\langle 1\rangle_0}.\\
\end{split}
\end{equation}
Next, we plug Eq.~(\ref{delmu1finale}) in Eq.~(\ref{simpliWSMBKEB1E0}), thereby eliminating one of the unknowns.
We still need to eliminate ${\bf E}_1$.
The strategy to follow is to take the average of Eq.~(\ref{simpliWSMBKEB1E0}) over the Fermi surface surrounding the node $\alpha$, and then
apply the normalization condition $\langle \chi_1^{(+)}\rangle_0 +\langle \chi_1^{(-)}\rangle_0 = 0$.
This gives the following expression for the longitudinal part of ${\bf E}_1$:
\begin{equation}\label{E1}
\begin{split}
    E_{1,\parallel}  = \frac{e}{\omega q\epsilon_{lat}}\bigg[i\textbf{q}\cdot\sum_{\alpha=+,-} \alpha\bigg(\frac{e}{\hbar}\frac{\langle\boldsymbol{\Omega}^{(\alpha)}\cdot\textbf{v}^{(\alpha)}\rangle_0\textbf{B}}{\langle1^{(\alpha)}\rangle_0} - \frac{\langle\partial_{\textbf{p}}(\textbf{m}^{(\alpha)}\cdot\textbf{B})\rangle_0}{\langle1^{(\alpha)}\rangle_0}\bigg)\bigg(1-\frac{\Gamma_{E}}{\Gamma_{A}}\bigg)\langle\chi_{0}^{(+)}\rangle_0 \\
    - i\textbf{q}\cdot\sum_{\alpha=+,-}\left(\frac{e}{\hbar}\langle(\boldsymbol{\Omega}^{(\alpha)}\cdot\textbf{v}^{(\alpha)})\lambda_{ij}^{(\alpha)}\rangle_0\textbf{B} - \langle\partial_{\textbf{p}}(\textbf{m}^{(\alpha)}\cdot \textbf{B})\lambda_{ij}^{(\alpha)}\rangle_0\right)\frac{\omega q_{j}u_{i}}{\Gamma_{A}}\bigg],\\
\end{split}
\end{equation}
where we have used $\langle\chi_{0}^{(+)}\rangle_0= -\langle\chi_{0}^{(-)}\rangle_0$, $\langle {\bf v}^{(\alpha)}\rangle_0=0$, $\langle ({\bf v}^{(\alpha)}\times{\bf B})\cdot\partial_{\bf p} \chi_0^{(\alpha)}\rangle_0=0$, $\sum_\alpha \langle \Omega^{(\alpha)}_i v_j^{(\alpha)}\rangle_0 =0$ (for any $i$ and $j$) and $\sum_\alpha \langle\partial_{\bf p}({\bf m}^{(\alpha)}\cdot{\bf B})\rangle_0=0$.
%The two latter equalities hold even for chiral WSM.
In addition, in Eq.~(\ref{E1}) we have omitted terms that are proportional to the free electron mass $m$.
We have verified that the contribution of the latter to the phonon magnetochiral effect is intrinsic (i.e., independent of $\Gamma_A$ and $\Gamma_E$) and geometric (i.e., proportional to $|C| $), but quantitatively negligible.

We note that $E_{1,\parallel}$ vanishes if the deformation potential has the same value in nodes of opposite chirality. Likewise, in the limit of $\Gamma_{A}\rightarrow\infty$, $E_{1,\parallel}$ vanishes if ${\bf q}\perp{\bf B}$ (this condition $\textbf{q}\cdot\textbf{B} =0$ holds true for spherical symmetry).
Interestingly, $E_{1,\parallel}$ is of purely geometrical origin (proportional to $|C|$).

Much like in the $B=0$ case, we will neglect the transverse component of ${\bf E}_1$. Then, using Eq. (\ref{E1}), we get 
\begin{equation}\label{totalavchi}
\begin{split}
&  \langle\chi_{1}^{(\alpha)}\rangle_0 =\frac{3e}{\hbar}\frac{\langle(\textbf{B}\cdot\boldsymbol{\Omega}^{(\alpha)})(\textbf{q}\cdot\textbf{v}^{(\alpha)})\rangle_0}{\Gamma_{E}}\frac{\langle\lambda_{ij}\rangle_0}{\langle1\rangle_0}q_{j}u_{i} -\frac{2e}{\hbar}\frac{\langle\textbf{v}^{(\alpha)}\cdot\boldsymbol{\Omega}^{(\alpha)}\rangle_0(\textbf{B}\cdot\textbf{q})}{\Gamma_{E}}\frac{\langle\lambda_{ij}\rangle_0}{\langle1\rangle_0} q_{j}u_{i}\\
&  -\frac{e}{\hbar}\langle(\textbf{B}\cdot\boldsymbol{\Omega}^{(\alpha)})(\textbf{q}\cdot\textbf{v}^{(\alpha)})\rangle_0\frac{i\omega}{\Gamma_{A}\Gamma_{E}}\frac{\langle\lambda_{ij}\rangle_0}{\langle1\rangle_0}q_{j}u_{i} - i\textbf{q}\cdot\bigg(\frac{e}{\hbar}\langle\boldsymbol{\Omega}^{(\alpha)}\cdot \textbf{v}^{(\alpha)}\rangle_0\textbf{B} -\langle\partial_{\textbf{p}}(\textbf{m}^{(\alpha)}\cdot\textbf{B})\rangle_0\bigg)\frac{\langle\chi_{0}^{(\alpha)}\rangle_0}{\langle1^{(\alpha)}\rangle_0}\bigg(\frac{1}{\Gamma_{E}} - \frac{1}{\Gamma_{A}}\bigg)\\
&  + i\textbf{q}\cdot\bigg(\frac{e}{\hbar}\langle(\boldsymbol{\Omega}^{(\alpha)}\cdot\textbf{v}^{(\alpha)})\lambda^{(\alpha)}_{ij}\rangle_0\textbf{B}
  -\langle\partial_{\textbf{p}}(\textbf{m}^{(\alpha)}\cdot\textbf{B})\lambda_{ij}^{(\alpha)}\rangle_0\bigg)\frac{\omega q_{j}u_{i}}{\Gamma_{E}\Gamma_{A}}- i\textbf{q}\cdot\bigg(\frac{e}{\hbar}\langle(\boldsymbol{\Omega}^{(\alpha)}\cdot\textbf{v}^{(\alpha)})\rangle_0\textbf{B}-\langle\partial_{\textbf{p}}(\textbf{m}^{(\alpha)}\cdot\textbf{B})\rangle_0\bigg)\frac{\langle\lambda_{ij}\rangle_0}{\langle1\rangle_0}\frac{\omega q_{j}u_{i}}{\Gamma_{E}\Gamma_{A}}\\
&  + \frac{\omega q\epsilon_{lat}}{\Gamma_{E} e}\bigg(\frac{e}{\omega q\epsilon_{lat}}\bigg)
  \frac{\langle1^{(\alpha)}\rangle_0}{\langle1\rangle_0}\bigg[i\textbf{q}\cdot\sum_{\beta=+,-} \beta \bigg(\frac{e}{\hbar}\frac{\langle\boldsymbol{\Omega}^{(\beta)}\cdot\textbf{v}^{(\beta)}\rangle_0\textbf{B}}{\langle1^{(\beta)}\rangle_0} - \frac{\langle\partial_{\textbf{p}}(\textbf{m}^{(\beta)}\cdot\textbf{B})\rangle_0}{\langle1^{(\beta)}\rangle_0}\bigg)\bigg(1-\frac{\Gamma_{E}}{\Gamma_{A}}\bigg)\langle\chi_{0}^{(+)}\rangle_0\\
    &- i\textbf{q}\cdot\sum_{\beta=+,-}\bigg\{\frac{e}{\hbar}\langle(\boldsymbol{\Omega}^{(\beta)}\cdot\textbf{v}^{(\beta)})\lambda_{ij}^{(\beta)}\rangle_0\textbf{B} - \langle\partial_{\textbf{p}}(\textbf{m}^{(\beta)}\cdot \textbf{B})\lambda_{ij}^{(\beta)}\rangle_0\bigg\}\frac{\omega q_{j}u_{i}}{\Gamma_{A}}\bigg].
\end{split}
\end{equation}
This equation satisfies the normalization condition $\langle\chi_{1}^{(+)}\rangle_0 + \langle\chi_{1}^{(-)}\rangle_0 =0$.
Much like $E_{1,\parallel}$,  $\langle\chi_1^{(\alpha)}\rangle_0$ is of geometrical origin and in the limit of $\Gamma_{A}\rightarrow\infty$, vanishes when ${\bf q}\perp{\bf B}$.
%The fact that a collinear ${\bf q}$ and ${\bf B}$ induce an electrochemical potential difference between the nodes is related to the chiral anomaly.

Equations (\ref{simpliWSMBKEB1E0}), (\ref{delmu1finale}), (\ref{E1})) and (\ref{totalavchi})), together with the results from the previous section and Eq.~(\ref{pert}), complete the solution of the BKE to first order in $B$.
This solution will enable us to derive the expressions for the sound velocity and attenuation.

\subsection*{V. Elasticity equations for lattice vibrations in the presence of Weyl fermions}
\label{sec:elasticity}
To calculate the velocity and attenuation of sound propagation, we use the elasticity equation for the lattice in the presence of conduction electrons
\begin{equation}\label{elasticity}
\rho \ddot{u}_{h} = \partial_{r_k}\sigma_{hk}^{\rm lat} +\bigg[\bigg( \textbf{j}_{\rm el}({\bf r},t) +  \textbf{j}_{\rm lat}({\bf r},t)\bigg)\times \textbf{B} +\textbf{F}({\bf r},t)\bigg]_{h} ,
\end{equation}
where $h\in\{x,y,z\}$,  $\rho$ is the mass density of the material, $\sigma^{\rm lat}$ is the stress tensor in the absence of conduction electrons, $\textbf{j}_{\rm el}$ is the electronic current density, $\textbf{j}_{\rm lat}$ is the ionic current density and $\textbf{F}$ is the drag force exerted by the electrons on the lattice.
In Eq.~(\ref{elasticity}), we have neglected a term involving the time derivative of the total electric current density.
This omission has been justified in the main text.

The stress tensor is related to strain through
\begin{equation}\label{stress}
\sigma_{hk}^{\rm lat} = s_{hkim} u_{im},
\end{equation}
where $s_{hkim}$ is the stiffness tensor whose general form depends on the crystal symmetry of the material.

The electronic current density is given as
%\cite{son}
\begin{equation}\label{eleccurrent}
\textbf{j}_{\rm el}({\bf r},t)= e\int\frac{d^{3}\textbf{p}}{(2\pi\hbar)^{3}}\left(1+ \frac{e}{\hbar}\textbf{B}\cdot\boldsymbol{\Omega}_\textbf{p}\right) \dot{\textbf{r}} f({\bf r},t),
\end{equation}
while the lattice current to first order in ${\bf u}$ reads 
\begin{equation}\label{latcurrent}
\textbf{j}_{\rm lat}({\bf r},t) = -n_{0}e\dot{\textbf{u}},
\end{equation}
$en_{0} = e\langle\langle f_{0}\rangle\rangle$ being the ionic charge (to zeroth order in $u$).
In linear response to the external magnetic field, it is sufficient to evaluate the total current density at zero field.
The outcome reads
\begin{equation}\label{totalcurrent}
\textbf{j} = \textbf{j}_{\rm el} + \textbf{j}_{\rm lat}= -e\langle \textbf{v}\chi_{0}\rangle_0 - \frac{e}{\hbar}\partial_{\textbf{r}}\times\dot{\textbf{u}}\langle\!\langle\Omega_i(p_i - mv_i)f_0\rangle\!\rangle_0.
\end{equation}
In the absence of the Berry curvature, the expression for ${\bf j}$ reduces to that shown in Ref. [\onlinecite{kontorovich1963}].
The part of the current coming from the Berry curvature is special in that it depends on all occupied electronic states and not just those at the Fermi surface.
One can show that $\langle\langle \W_{\bf p}\cdot{\bf v} f_0\rangle\rangle_0=0$ because the contributions from the two nodes of opposite chirality cancel (regardless of the crystal being chiral or not). In contrast, $\langle\langle \W_{\bf p}\cdot{\bf p} f_0\rangle\rangle_0\neq 0$. According to our estimates, this term is nonetheless small and its impact in the phonon magnetochiral effect will turn out to be quantitatively unimportant.

Therefore, the Lorentz force acting on the current is
\begin{equation}\label{alpherrub}
\begin{split}
\textbf{j}\times\textbf{B} & = \bigg[-e\langle \textbf{v}\chi_{0}\rangle_0 - \frac{e}{\hbar}\partial_{\textbf{r}}\times\dot{\textbf{u}}\langle\!\langle\Omega_i(p_i - mv_i)f_0\rangle\!\rangle_0\bigg]\times \textbf{B}\\
\quad &\simeq \frac{e}{\Gamma_{A}}\langle (\textbf{v}\times\textbf{B})(\textbf{q}\cdot\textbf{v})\rangle_0\frac{\langle\lambda_{ij}\rangle_0}{\langle1\rangle_0}q_{j}u_{i} -\frac{e}{\hbar}\left((\textbf{B}\cdot\partial_{\bf r})\dot{\bf u} - \partial_{\bf r}(\dot{\bf u}\cdot\textbf{B})\right)\langle\!\langle\Omega_i(p_i - mv_i)f_0\rangle\!\rangle_0. \\
\end{split}
\end{equation}

To first order in $B$, the $h$ component of the drag force is given as
\begin{equation}\label{dragforce}
  F_h({\bf r},t) = \partial_{r_k} \langle\!\langle \lambda_{hk} f\rangle\!\rangle_{1} = \partial_{r_k}\int\frac{d^{3}\textbf{p}}{(2\pi\hbar)^{3}}\left(1+ \frac{e}{\hbar}\textbf{B}\cdot\boldsymbol{\Omega}_\textbf{p}\right)\lambda_{hk} f({\bf r},t).  
\end{equation}
%The second term in the rhs of Eq.~(\ref{elasticity}) is the Alpher-Rubin term which comprises of the contribution from the total current $\textbf{j}$ (given by Eq.~(\ref{totalcurrent})) and magnetic field. We calculate this term with total current density $\textbf{j}(B=0)$

In sum, in order to compute the right hand side of Eq.~(\ref{elasticity}) to first order in $B$, we require the knowledge of the electronic distribution function $f$ to the same order order in $B$.
Following Eq.~(\ref{f}),
\begin{equation}\label{dragf1}
f \approx f_{0}(\varepsilon^{(0)}_{\bf p}) + (\lambda_{ij}u_{ij} +p_i \tilde{v}_j u_{i j} - m \tilde{v}_i \dot{u}_i-\delta\mu+\chi)\frac{\partial  f_{0}}{\partial  \varepsilon_{\bf p}^{(0)}}.
\end{equation}
The first term in the right hand side of Eq.~(\ref{dragf1}) does not depend on space (it is the equilibrium distribution with chemical potential $\mu_0$) and thus it does not contribute to the drag force.
The remaining terms lead to
\begin{align}
  \label{drag2}
  F_h &= -\partial_{r_k}\langle \lambda_{h k}\left(\lambda_{i j} u_{i j} + p_i \tilde{v}_j u_{i j} - m \tilde{v}_i\dot{u}_i - \delta\mu_0-\delta\mu_1 + \chi_0+\chi_1\right) \rangle_1\nonumber\\
  &\simeq -\partial_{r_k}\langle \lambda_{h k} \chi_1\rangle_0 +(\partial_{r_k}\delta\mu_1)\langle\lambda_{h k}\rangle_0 -\partial_{r_k} \left\langle\frac{1}{v_F} ({\bf m}_{{\bf p}_F}\cdot{\bf B}) \partial_p (\lambda_{h k}\chi_0)\right\rangle_0
    +\text{ terms independent of $B$},
\end{align}
where we have neglected an unimportant term proportional to the bare electron mass $m$.
We do not explicitly write the zeroth order terms in the magnetic field, because we will be interested in predicting only the $B$-dependence of the sound velocity and attenuation.
The third term in Eq.~(\ref{drag2}) is proportional to the derivatives of the deformation potentials $\lambda_1$ and $\lambda_2$ with respect to the energy.
We will hereafter ignore these terms (by implicitly assuming that $\lambda_1$ and $\lambda_2$ depend weakly on energy in the vicinity of the Fermi surface).
%Even if we did not neglect $\partial\lambda_{1(2)}/\partial\epsilon_0$, the third term in Eq.~(\ref{drag2}) would still be smaller than the ones we keep

Upon Fourier transforming, we get
%Here the terms independent of $u$ do not contribute to the drag force because they vanish upon taking spatial derivative. Re-writing the above equation in terms of Fermi surface averages
\begin{equation}\label{drag3}
\begin{split}
  F_h\simeq&-i q_k\langle \lambda_{h k} \chi_1\rangle_0 +i q_k \delta\mu_1\langle\lambda_{h k}\rangle_0 +\text{ terms independent of $B$}.
 \end{split}
\end{equation}
In order to connect with the results from the previous section, Eq.~(\ref{drag3}) may be rewritten as
\begin{equation}
\label{drag4}
  F_h\simeq-i q_k \sum_{\alpha=+,-}\left(\langle\lambda^{(\alpha)}_{h k} \chi^{(\alpha)}_1\rangle_0 - \delta\mu_1\langle\lambda^{(\alpha)}_{h k}\rangle_0\right) +\text{ terms independent of $B$}.
\end{equation}

%Next, we write the eigen value equation corresponding to the elasticity equation given by Eq.~(\ref{elasticity}). We will solve the eigenvalue equation to derive the velocity and attenuation of sound propagation.

\subsection*{VI. Velocity and attenuation of sound waves in WSM: phonon magnetochiral effect}

Inserting the expressions for the drag force (Eq.~(\ref{drag3})) and the Lorentz force (Eq.~(\ref{alpherrub})) in Eq.~(\ref{elasticity}), we rewrite the elasticity equation as
\begin{equation}\label{eigenscalar}
\begin{split}
  \rho\omega^2 u_h &=s_{hkim} q_{k}q_{m} u_{i}-\frac{e}{\Gamma_{A}}\langle (\textbf{v}\times\textbf{B})_{h}(\textbf{q}\cdot\textbf{v})\rangle_0\frac{\langle\lambda_{ij}\rangle_0}{\langle1\rangle_0}q_{j}u_{i} +\frac{e}{\hbar}\left((\textbf{B}\cdot\textbf{q})\omega u_{i}\delta_{hi} - q_{h}(\omega u_{i}B_{i})\right)\langle\langle\Omega_j(p_j - mv_j)f_0\rangle\rangle_0\\
  &-iq_{k}\delta\mu_1\langle\lambda_{hk}\rangle_0 + i q_k \langle\lambda_{hk}\chi_{1}\rangle_0,\\
\end{split}
\end{equation}
where $h$ is a fixed index and the rest are summed over.
Thus Eq. (\ref{eigenscalar}) is a system of three equations (one for each value of $h$) that can be written in matrix form and solved as an eigenvalue problem.
The eigenvalues give $\omega$ as a function of ${\bf q}$ and ${\bf B}$, while the eigenvectors give the direction of ${\bf u}$ for the three acoustic modes.
The terms of the drag force that are independent of the magnetic field renormalize the phonon frequency and render a nonzero linewidth; these effects will be implicitly absorbed through a redefinition of the stiffness tensor and by adding an imaginary $B$-independent part to the phonon frequency in Eq.~(\ref{eigenscalar}).

From here on we fix ${\bf B}=B_z\hat{\bf z}$, where $\hat{\bf z}$ is a high symmetry direction of a chiral cubic crystal (point group $O$ or $T$).
We first consider the configuration in which sound propagates along $\hat{\bf z}$, with a wave vector ${\bf q}=q_z\hat{\bf z}$.
Both $B_z$ and $q_z$ may be either positive or negative.
For this configuration, Eq. (\ref{eigenscalar}) becomes

\begin{align}
  \label{Mxyz}
\rho \omega^2 u_z &= q_z^2 s_{zzzz} u_z - i q_z \delta\mu_1 \langle\lambda_{zz}\rangle_0 + i q_z \langle \lambda_{zz}\chi_1\rangle_0\nonumber\\
\rho \omega^2 u_x &= q_z^2 s_{xzxz} u_x + \frac{e}{\hbar} B_z q_z \omega u_x \langle\langle \Omega_j (p_j - m v_j) f_0\rangle\rangle_0 + i q_z\langle \lambda_{zx} \chi_1\rangle_0\nonumber \\
\rho \omega^2 u_y &= q_z^2 s_{yzyz} u_y + \frac{e}{\hbar} B_z q_z \omega u_y \langle\langle \Omega_j (p_j - m v_j) f_0\rangle\rangle_0 + i q_z\langle \lambda_{zy} \chi_1\rangle_0.
\end{align}
One can readily show that $\delta\mu_1$ and $\langle \lambda_{zz} \chi_1\rangle_0$ involve only $u_z$.
Likewise, it can be shown that the second and third lines in Eq.~(\ref{Mxyz}) contain $u_x$ and $u_y$, but not $u_z$.
Accordingly, the first line of Eq.~(\ref{Mxyz}) describes a longitudinal acoustic phonon.
The second and third line correspond to transverse acoustic phonons.

%and that $\langle\lambda_{z j}\chi_1\rangle_0$ only involves $u_j$.

%For the cubic symmetry we consider, the two transverse modes have identical dispersion.

Let us first determine the dispersion of the longitudinal mode.
On one hand, we need
\begin{equation}
\label{2nodeft}
i q_z \delta\mu_1 \langle \lambda_{zz}\rangle_0 = - q_z^2 \frac{\langle\lambda_{zz}\rangle_0}{\langle 1\rangle_0} \frac{\epsilon_{lat}}{e} E_{1,\parallel},
\end{equation}
where we have used Eq.~(\ref{delmu1finale}) and have neglected an unimportant term proportional to the electron mass $m$.
Plugging the value of electric field $E_{1,\parallel}$ from Eq.~(\ref{E1}), we reach
\begin{equation}\label{inte2}
\begin{split}
  i q_z \delta\mu_1 \langle\lambda_{zz}\rangle_0\simeq i u_z q_{z}^{3}\frac{eB_{z}|C|}{3\pi^{2}\hbar^{2}}\frac{1}{\Gamma_{E}}\frac{\langle\lambda_{zz}\rangle_0}{\langle1\rangle_0}\bigg(\frac{\langle\lambda^{(+)}_{zz}\rangle_0}{\langle1^{(+)}\rangle_0}-\frac{\langle\lambda_{zz}^{(-)}\rangle_0}{\langle1^{(-)}\rangle_0}\bigg),\\
\end{split}
\end{equation}
where we have used the expression of $\langle\chi_{0}^{(+)}\rangle$ from Eq.~(\ref{avchi0}) and we have omitted terms that are a factor $\Gamma_A/\Gamma_E$ smaller than the terms shown.
This omission is justified insofar as $\Gamma_A\gg \Gamma_E$, which is believed to hold in WSM.
In addition, in the derivation of Eq.~(\ref{inte2}), we have used the relations
\begin{align}
\label{eq:relations}
  \langle\W^{(\alpha)}\cdot{\bf v}^{(\alpha)}\rangle_0 &= \frac{\alpha |C|}{4 \pi^2\hbar}\nonumber\\
  \langle \partial_{p_z} m_z^{(\alpha)}\rangle_0 &= -\frac{e \alpha |C|}{12 \pi^2\hbar^2}.
\end{align}

On the other hand, resorting to similar approximations (e. g., invoking $\Gamma_A\gg \Gamma_E$), we obtain
\begin{align}
  \label{inte3}
  -i q_z\langle\lambda_{zz}\chi_1\rangle_0  &\simeq i u_z q_{z}^{3}\frac{eB_{z}|C|}{4\pi^{2}\hbar^{2}}\frac{1}{\Gamma_{E}}\frac{\langle\lambda_{zz}\rangle_0}{\langle1\rangle_0}\bigg(\frac{\langle\lambda^{(+)}_{zz}\rangle_0}{\langle1^{(+)}\rangle_0}-\frac{\langle\lambda_{zz}^{(-)}\rangle_0}{\langle1^{(-)}\rangle_0}\bigg)\nonumber\\
  &-u_z q_z^3 \frac{e B_z |C|}{3\pi^2 \hbar^2}\frac{\omega}{\Gamma_E^2}\frac{\langle 1^{(+)}\rangle_0-\langle 1^{(-)}\rangle_0}{\langle 1^{(+)}\rangle_0+\langle 1^{(-)}\rangle_0} \bigg(\frac{\langle\lambda^{(+)}_{zz}\rangle_0}{\langle1^{(+)}\rangle_0}-\frac{\langle\lambda^{(-)}_{zz}\rangle_0}{\langle1^{(-)}\rangle_0}\bigg)^2.
  \end{align}
%{\bf danger of cancellation}
Inserting Eq.~(\ref{inte2}) and Eq.~(\ref{inte3}) in the first line of Eq.~(\ref{Mxyz}), we have
\begin{equation}\label{longmode}
\begin{split}
  0 &= q_{z}^{2}s_{33} -\rho\omega^{2} - i q_{z}^{3}\frac{7 e B_{z}|C|}{12\pi^{2}\hbar^{2}}\frac{1}{\Gamma_{E}}\frac{\langle\lambda_{zz}\rangle_0}{\langle1\rangle_0}\bigg(\frac{\langle\lambda^{(+)}_{zz}\rangle_0}{\langle1^{(+)}\rangle_0}-\frac{\langle\lambda^{(-)}_{zz}\rangle_0}{\langle1^{(-)}\rangle_0}\bigg)\\
  &+ q_z^3 \frac{e B_z |C|}{3\pi^2 \hbar^2}\frac{\omega}{\Gamma_E^2}\frac{\langle 1^{(+)}\rangle_0-\langle 1^{(-)}\rangle_0}{\langle 1^{(+)}\rangle_0+\langle 1^{(-)}\rangle_0} \bigg(\frac{\langle\lambda_{zz}^{(+)}\rangle_0}{\langle1^{(+)}\rangle_0}-\frac{\langle\lambda_{zz}^{(-)}\rangle_0}{\langle1^{(-)}\rangle_0}\bigg)^2,
\end{split}
\end{equation}
where $s_{33}=s_{zzzz}$.
The magnetic-field corrections to the acoustic phonon dispersion are odd in $q_z$ as well as in $B_z$, and proportional to $|C|$.
This confirms the existence of a phonon magnetochiral effect of purely band-geometric origin.

Solving Eq.~(\ref{longmode}), we get
\begin{align}
\label{eq:solong}
  \omega &\simeq c_s(0) |q_z| + i\gamma(0) + \delta\omega_R + i \delta\omega_I,\nonumber\\
  \delta\omega_R &\simeq q_z^3 \frac{e B_z |C|}{6\pi^2 \hbar^2\rho}\frac{1}{\Gamma_E^2}\frac{\langle 1^{(+)}\rangle_0-\langle 1^{(-)}\rangle_0}{\langle 1^{(+)}\rangle_0+\langle 1^{(-)}\rangle_0} \bigg(\frac{\langle\lambda^{(+)}_{zz}\rangle_0}{\langle1^{(+)}\rangle_0}-\frac{\langle\lambda^{(-)}_{zz}\rangle_0}{\langle1^{(-)}\rangle_0}\bigg)^2\nonumber,\\
  \delta\omega_I &\simeq -q_{z} |q_z|\frac{7 e B_{z}|C|}{24\pi^{2}\hbar^{2}\rho c_s(0)}\frac{1}{\Gamma_{E}}\frac{\langle\lambda_{zz}\rangle_0}{\langle1\rangle_0}\bigg(\frac{\langle\lambda^{(+)}_{zz}\rangle_0}{\langle1^{(+)}\rangle_0}-\frac{\langle\lambda^{(-)}_{zz}\rangle_0}{\langle1^{(-)}\rangle_0}\bigg),
\end{align}
where $c_s(0)=\sqrt{s_{33}/\rho}$ and $\gamma(0)$ are the sound velocity and attenuation in the absence of magnetic field.
The latter has been added by hand to describe damping of the sound waves at $B=0$.
We note that, for a chiral WSM in which $\lambda_{1(2)}^{(+)}+\lambda_{1(2)}^{(-)}$ is of the same order as $\lambda_{1(2)}^{(+)}-\lambda_{1(2)}^{(-)}$, $\delta\omega_R/\delta\omega_I \simeq c_s(0) q_z/\Gamma_E\ll 1$.
Thus, we anticipate that the magnetochiral effect will be more pronounced in the sound attenuation than in the sound velocity.
%Strictly speaking, $c_s(0)$ should have a small imaginary part describing attenuation in the absence of a magnetic field. 

The correction to the sound velocity due to the magnetic field is given by
\begin{equation}\label{velocity}
\begin{split}
\delta c_{s} &= \frac{\partial \delta\omega_{R}}{\partial |q_{z}|}
= q_z |q_z| \frac{e B_z |C|}{2\pi^2 \hbar^2\rho}\frac{1}{\Gamma_E^2}\frac{\langle 1^{(+)}\rangle_0-\langle 1^{(-)}\rangle_0}{\langle 1^{(+)}\rangle_0+\langle 1^{(-)}\rangle_0} \bigg(\lambda_1^{(+)}-\lambda_1^{(-)} + \frac{\lambda_2^{(+)}}{3}-\frac{\lambda_2^{(-)}}{3}\bigg)^2,
\end{split}
\end{equation}
where we have used the relation
\begin{equation}
\label{relation2}
  \frac{\langle\lambda^{(+)}_{zz}\rangle_0}{\langle1^{(+)}\rangle_0}-\frac{\langle\lambda^{(-)}_{zz}\rangle_0}{\langle1^{(-)}\rangle_0}=\lambda_1^{(+)}-\lambda_1^{(-)} + \frac{\lambda_2^{(+)}}{3}-\frac{\lambda_2^{(-)}}{3}.
\end{equation}
Therefore, the phonon magnetochiral effect in the sound velocity is given as
\begin{equation}\label{vmc}
\begin{split}
v_{\rm{MC}}&= \frac{c_{s}(\textbf{B}\parallel\hat{\textbf{q}})-c_{s}(\textbf{B}\parallel-\hat{\textbf{q}})}{c_{s}(0)}\\
& \approx \frac{e q_{z} |q_z|B_{z}|C|}{\pi^2\hbar^2\rho c_{s}(0)}\frac{1}{\Gamma_{E}^{2}}\frac{\langle1^{(+)}\rangle_0 - \langle1^{(-)}\rangle_0}{\langle1^{(+)}\rangle+\langle1^{(-)}\rangle}\bigg(\lambda_1^{(+)}-\lambda_1^{(-)} + \frac{\lambda_2^{(+)}}{3}-\frac{\lambda_2^{(-)}}{3}\bigg)^{2}.
\end{split}
\end{equation}
Clearly, broken inversion and mirror symmetries are required in order to have $v_{\rm{MC}}\neq 0$.
In order to proceed with a numerical estimate, we write
\begin{equation}
  \frac{\langle1^{(+)}\rangle - \langle1^{(-)}\rangle}{\langle1^{(+)}\rangle+\langle1^{(-)}\rangle} = \frac{1-\left(\frac{\varepsilon_F^{(-)}}{\varepsilon_F^{(+)}}\right)^2 \left(\frac{v_F^{(+)}}{v_F^{(-)}}\right)^3}{1+\left(\frac{\varepsilon_F^{(-)}}{\varepsilon_F^{(+)}}\right)^2 \left(\frac{v_F^{(+)}}{v_F^{(-)}}\right)^3},
  \end{equation}
where $\varepsilon_F^{(\alpha)}=v_F^{(\alpha)} p_F^{(\alpha)}$ is the distance in energy from the Weyl node $\alpha$ to the equilibrium chemical potential.
Using $\varepsilon_F^{(+)}=20\,{\rm meV}$, $\varepsilon_F^{(-)}= 5\,{\rm meV}$, $v_F^{(+)}=10^5\, {\rm m/s}$, $v_F^{(-)}=1.5\times10^5\,{\rm m/s}$, $\rho=10^4\,{\rm kg/m}^3$, $B_z=1\,{\rm T}$, $c_s(0)=2\times10^3\,{\rm m/s}$, $q=0.5\times10^6 \,{\rm m^{-1}}$,   $\lambda_{1(2)}= 1\sim 2$ eV, we get $v_{\rm{MC}}$ of the order of a few parts per million.
This is a priori measurable in state-of-the-art ultrasound velocity measurements, whose resolution is about one p.p.m.
%For $\rho = 10^{4}Kg/m^{3}$, magnetic field $B=1$T, $E_{F}^{+} = 10 $meV $\sim 20$ meV, $E_{F}^{-} = 5 $meV, $v_{F}^{+} = 10^{5} m/s$, $v_{F}^{-} =1.5\times 10^{5} m/s $, the value of 
%\begin{equation}\label{dos}
%\bigg(\frac{\langle1\rangle^{(+)} - \langle1\rangle^{(-)}}{\langle1\rangle^{(+)}+\langle1\rangle^{(-)}}\bigg) = 0.86 \sim 0.96,
%\end{equation}
%therefore, as the difference between $E_{F}^{+}$ and $E_{F}^{-}$ increases, the ratio in Eq.~(\ref{dos}) $\approx 1$. For the chiral Weyl semimetals, the difference in Fermi energies bet%ween the nodes is generally quite large \cite{chang, tang} and hence, Eq.~(\ref{dos}) is validated. Therefore, we write Eq.~(\ref{vmc}) as
%\begin{equation}\label{vmc}
%v_{\rm{MC}}\approx \frac{4}{\rho}\frac{eq_{z}^{2} sg(q_{z})B_{z}|C|}{c_{s}(0)}\bigg(\frac{1}{\pi\hbar\Gamma_{E}}\bigg)^{2}\bigg(\lambda_{1,ps} +\frac{\lambda_{2,ps}}{3}\bigg)^{2}.
%\end{equation}
%The origin of the above term can be traced back to the contribution from $\langle\chi_{0}^{(\alpha)}\rangle$ in Eq.~(\ref{simpliinte3}). The term $\langle\chi_{0}^{(\alpha)}\rangle$ is non-zero only if the deformation potential is different on the two Weyl nodes. For phonon wave vector $q = 10^{-6} m^{-1}$, the phonon magnetochiral effect in sound velocity is $v_{\rm{MC}} \approx 7 \times 10^{-6}$.

Concerning the magnetochiral effect in the sound attenuation, it can be characterized by the dimensionless ratio
%Similarly from Eq.~(\ref{longmode}), the imaginary part of the frequency is given as
%\begin{equation}\label{wi}
%\omega_{I}(\textbf{q},\textbf{B}) \approx  -\frac{7}{12}\frac{q_{z}^{2}sg(q_{z})}{\rho c_{s}(0)}\frac{eB_{z}}{\pi^{2} \hbar^{2}}\frac{|c|}{\Gamma_{E}}\frac{\langle\lambda_{zz}\rangle}{\langle1\rangle}\bigg(\lambda_{1,ps}+\frac{\lambda_{2,ps}}{3}\bigg).
%\end{equation}
%Attenuation of the sound is quantified by factor $r_{\rm{MC}}$ which can be written as
\begin{equation}\label{rmc}
  r_{\rm{MC}} \approx \frac{2L \delta\omega_{I}}{c_{s}(0)},
\end{equation}
where $L$ is the thickness of the sample traversed by the sound wave.
The contribution from $\delta c_s$ to $r_{\rm{MC}}$ has been omitted from Eq.~(\ref{rmc}) because it is relatively negligible when $\delta\omega_I\gg\delta\omega_R$ and $\gamma(0)\ll c_s(0) q_z$.
The ratio $r_{\rm{MC}}$  describes the relative change in the decay of the amplitude of the sound wave traversing the sample when the propagation direction is parallel and antiparallel to the magnetic field.
For the numerical parameters presented above (in addition to $L=1 \,{\rm cm}$), Eqs. (\ref{eq:solong}) and (\ref{rmc}) yield $r_{\rm{MC}} \approx 0.31$, which is large and a priori easily detectable.

Having completed the discussion about the longitudinal mode, let us next investigate the transverse modes. From Eqs. (\ref{Mxyz}) and (\ref{simpliWSMBKEB1E0}), we get
\begin{equation}\label{Gtransxy}
\begin{split}
0 = (q_{z}^{2}s_{xzxz} -\rho\omega^{2})u_{x} + \frac{e}{\hbar}(B_{z}q_{z}\omega u_{x})\langle\langle\Omega_{j}(p_{j}-mv_{j})f_{0}\rangle\rangle_{0} + iq_{z}\bigg[\bigg\langle \lambda_{xz}i(\textbf{q}\cdot\textbf{B})\frac{e}{\hbar}(\boldsymbol{\Omega}\cdot\textbf{v})\frac{\lambda_{xz}q_{z}\omega u_{x}}{\Gamma_{A}^{2}}\bigg\rangle_{0}\\   
+\bigg\langle\lambda_{xz}i(\textbf{q}\cdot\textbf{B})\frac{ev_{F}\hbar\alpha|C|}{2p^{2}}(1-2\cos^{2}\theta)\frac{\lambda_{xz}\omega q_{z}u_{x}}{\Gamma_{A}^{2}}\bigg\rangle_{0} + \bigg\langle\lambda_{xz} e v_{x}B_{z}\frac{2p_{z}}{p^{2}}\frac{\omega q_{z} u_{y}}{\Gamma_{A}^{2}}\bigg\rangle_{0}\bigg]\\
0 = (q_{z}^{2}s_{yzyz} -\rho\omega^{2})u_{y} + \frac{e}{\hbar}(B_{z}q_{z}\omega u_{y})\langle\langle\Omega_{j}(p_{j}-mv_{j})f_{0}\rangle\rangle_{0} + iq_{z}\bigg[\bigg\langle \lambda_{yz}i(\textbf{q}\cdot\textbf{B})\frac{e}{\hbar}(\boldsymbol{\Omega}\cdot\textbf{v})\frac{\lambda_{yz}q_{z}\omega u_{y}}{\Gamma_{A}^{2}}\bigg\rangle_{0}\\   
+\bigg\langle\lambda_{yz}i(\textbf{q}\cdot\textbf{B})\frac{ev_{F}\hbar\alpha|C|}{2p^{2}}(1-2\cos^{2}\theta)\frac{\lambda_{yz}\omega q_{z}u_{y}}{\Gamma_{A}^{2}}\bigg\rangle_{0} + \bigg\langle\lambda_{yz} e v_{y}B_{z}\frac{2p_{z}}{p^{2}}\frac{\omega q_{z} u_{x}}{\Gamma_{A}^{2}}\bigg\rangle_{0}\bigg].\\
\end{split}
\end{equation}
This set of equations can be rearranged in the form
\begin{equation}
  \left(\begin{array}{cc} a+i b & 0 \\
    0 & a-i b
    \end{array}\right)\left(\begin{array}{c} u_x+ i u_y \\ u_x-i u_y\end{array}\right) = \left(\begin{array}{c} 0 \\ 0\end{array}\right),
\end{equation}
where
\begin{align}
  a&=\left(q_{z}^{2} s_{xzxz} -\rho\omega^{2}\right) + \frac{e}{\hbar}(B_{z}q_{z}\omega)\langle\langle\Omega_{j}(p_{j}-mv_{j})f_{0}\rangle\rangle_{0}- \frac{e}{\hbar}\frac{q_{z}^{3}B_{z}\omega}{\Gamma_{A}^{2}}\left\langle \lambda_{xz}^{2}(\boldsymbol{\Omega}\cdot\textbf{v})\right\rangle_{0}  -\frac{q_{z}^{3}B_{z}\omega\alpha|C|e\hbar}{2\Gamma_{A}^{2}}\left\langle\frac{\lambda_{xz}^{2}(1-\cos^{2}\theta)}{p^{2}}\right\rangle_{0}\nonumber\\
b &= -\frac{2ieq_{z}^{2}B_{z}\omega}{\Gamma_{A}^{2}}\left\langle\frac{\lambda_{xz}v_{x}p_{z}}{p^{2}}\right\rangle_{0}.
\end{align}
Here, we have used $s_{xzxz}=s_{yzyz}$ and have exploited the cubic symmetry of the angular integrals (e.g. $\langle\lambda_{x z} v_x v_z\rangle_0=\langle \lambda_{y z} v_y v_z\rangle_0$).
Thus, we learn that the transverse modes are circularly polarized in the presence of a magnetic field.

One important aspect of Eq.~(\ref{Gtransxy}) is that it does not contain $\Gamma_E$.
The reason is that transverse modes do not produce any chiral charge imbalance (Eqs.~(\ref{avchi0}) and (\ref{totalavchi}) both give zero).
Solving the equations $a+i b=0$ and $a-ib=0$, we get the dispersion relations for the two transverse modes.
These solutions do display a phonon magnetochiral effect.
Yet, the effect is quantitatively negligible compared to the magnetochiral effect for the longitudinal mode.
The latter is made larger by the fact that the chiral charge imbalance produced by longitudinal phonons relaxes slowly.

Finally let us look at the configuration $\textbf{q}\perp\textbf{B}$. Plugging $\textbf{q} = q_x\hat{\textbf{x}}$ and $\textbf{B} = B_z\hat{\textbf{z}}$ in Eq.~(\ref{eigenscalar}), we have
\begin{equation}\label{configurationperp}
\begin{split}
\rho\omega^{2}u_{z} &= s_{zxzx}q_{x}^{2}u_{z} + iq_{x}\langle\lambda_{zx}\chi_{1}\rangle_{0}\\
\rho\omega^{2}u_{x} &= s_{xxxx}q_{x}^{2}u_{x} - iq_{x}\delta\mu_{1}\langle\lambda_{xx}\rangle_{0} + iq_{x}\langle\lambda_{xx}\chi_{1}\rangle_{0}\\
\rho\omega^{2}u_{y} &= s_{yxyx}q_{x}^{2}u_{y} + iq_{x}\langle\lambda_{yx}\chi_{1}\rangle_{0}.\\
\end{split}
\end{equation}
Using the expression of $\chi_{1}$ from Eq.~(\ref{simpliWSMBKEB1E0}), the first equation in Eq.~(\ref{configurationperp}) takes the form

\begin{equation}\label{uzqperpB}
\begin{split}
\rho\omega^{2}u_{z} &= s_{zxzx}q_{x}^{2}u_{z} + \bigg[i\frac{q_{x}^{3}}{\Gamma_{A}}\frac{\langle\lambda_{xx}\rangle_{0}}{\langle1\rangle_{0}}\langle\lambda_{zx}\frac{e}{\hbar}(B_{z}\Omega_{z})v_{x}\rangle_{0} + \frac{\omega}{\Gamma_{A}^{2}}q_{x}^{3}\frac{\langle\lambda_{xx}\rangle_{0}}{\langle1\rangle_{0}}\langle\lambda_{xz}\frac{e}{\hbar}B_{z}\Omega_{z}v_{x}\rangle_{0}\bigg] u_{x}.\\
\end{split}
\end{equation}

Let us now look at the second equation in Eq.~(\ref{configurationperp}). Here we will use the expressions for $\delta\mu_{1}$, $E_{1,\parallel}$ and $\langle\chi_{1}^{(\alpha)}\rangle$, such that 
\begin{equation}\label{uxqperpB}
\begin{split}
\rho\omega^{2}u_{x} &= s_{xxxx}q_{x}^{2}u_{x} +\bigg[i\frac{q_{x}^{3}}{\Gamma_{A}}\frac{\langle\lambda_{xx}\rangle_{0}}{\langle1\rangle_{0}}\langle\lambda_{xz}\partial_{p_x}({\bf m}_{\textbf{p}}\cdot\textbf{B})\rangle_{0} + \langle\lambda_{xx}\lambda_{xz} \partial_{p_x} ({\bf m}_{\textbf{p}}\cdot\textbf{B})\rangle_{0}\frac{\omega q_{x}^{3}}{\Gamma_{A}^{2}}\\
& -q_{x}^{3}\frac{\omega}{\Gamma_{A}^{2}}\sum_{\alpha =+,-}\frac{\langle\lambda_{xx}^{(\alpha)}\rangle}{\langle1\rangle_{0}}\langle\lambda_{xz}^{(\alpha)}\partial_{p_x} ({\bf m}^{(\alpha)}_{\textbf{p}}\cdot\textbf{B})\rangle_{0}-q_{x}^{3}\sum_{\alpha=+,-}\frac{\langle\lambda_{xx}^{(\alpha)}\rangle}{\langle1^{(\alpha)}\rangle_{0}}\frac{\omega}{\Gamma_{E}\Gamma_{A}}\langle\lambda_{zx}^{(\alpha)} \partial_{p_x}({\bf m}^{(\alpha)}_{\textbf{p}}\cdot\textbf{B})\rangle_{0}\\
& +q_{x}^{3}\frac{\omega}{\Gamma_{E}\Gamma_{A}}\sum_{\alpha =+,-}\frac{\langle\lambda_{xx}^{(\alpha)}\rangle_{0}}{\langle1\rangle_{0}}\langle\lambda_{zx}^{(\alpha)} \partial_{p_x}({\bf m}^{(\alpha)}_{\textbf{p}}\cdot\textbf{B})\rangle_{0}\bigg]u_{z} -iq_{x}^{2}\frac{eB_{z}\omega}{\Gamma_{A}^{2}} \bigg[\sum_{\alpha=+,-}\lambda_{2}^{(\alpha)}\bigg\langle\frac{\lambda_{xx}^{(\alpha)}p_{x}^{(\alpha)}v_{x}^{(\alpha)}}{p^{2}}\bigg\rangle_{0}\bigg]u_{y}.\\
\end{split}
\end{equation}

%{\bf Typos in the preceding equation?}

For the third equation in Eq.~(\ref{configurationperp}) we get
\begin{equation}\label{uyqperp}
\rho\omega^{2}u_{y} = s_{yxyx}q_{x}^{2}u_{y} - 2iq_{x}^{2}\frac{eB_{z} \omega}{\Gamma_{A}^{2}}\bigg[\sum_{\alpha=+,-}\lambda_{2}^{(\alpha)}\bigg\langle\frac{\lambda_{yx}^{(\alpha)}v_{y}^{(\alpha)}p_{x}^{(\alpha)}}{p^{2}}\bigg\rangle_{0}\bigg] u_{x}.
\end{equation}

The aforementioned three equations may be once again solved as an eigenvalue problem, leading to the dispersions of the three acoustic modes. 
Nevertheless, compared to the Eq.~(\ref{longmode}), the contribution of the magnetic field to the dispersions is smaller by a factor $\Gamma_A/\Gamma_E$. 
Thus, the phonon magnetochiral effect is unlikely to be measurable in the $\textbf{q}\perp\textbf{B}$ configuration. 

\subsection*{VII. Phonon magnetochiral effect (PMCE) in a Weyl semimetal model of 2$n$-nodes ($n>1$)}

Previously, we have focused on a minimal two-node model for WSM, in this section, we will provide generalization to the case of 2$n$ Weyl nodes (with $n >1$, relevant to Weyl semimetal with time-reversal symmetry). This generalization is relevant for the purposes of our work because we are mainly interested in the non-magnetic WSM. 

We begin by considering a time-reversal symmetric WSM with four Weyl nodes ($n=2$): 1, 2, 3 and 4.
We assume that 1 and 3 are partners under time reversal, while 2 and 4 are also partners under time reversal.
Nodes 1 and 3 have a chirality of $+1$, while nodes 2 and 4 have a chirality of $-1$.
Because of time-reversal symmetry, nodes 1 and 3 have the same Hamiltonian, hence the same electronic group velocities, same Berry curvatures, same orbital magnetic moments.
Likewise for the nodes 2 and 4. 

We will assume that the form of the collision term given by Eq.~(\ref{coll}) is still valid for the 4-node model. With this proviso, we start from Eq.~(\ref{E0chi0}), and write the expression for the solution of the BKE for the absence of magnetic field in this model as
\begin{align}
\label{eq:1}
  \chi_0^{(\alpha)} &= -\frac{\lambda_{i j}^{(\alpha)} \omega q_j u_i}{\Gamma_A} + \frac{\langle \lambda_{i j}\rangle_{0}}{\langle 1\rangle_0} \frac{\omega q_j u_i}{\Gamma_A} - \frac{m \omega^2 v_i^{(\alpha)} u_i}{\Gamma_A} + \frac{\omega \epsilon_{\rm lat} {\bf q}\cdot{\bf E}_0}{e \langle 1\rangle_0 \Gamma_A} - \frac{e {\bf E}_0\cdot{\bf v}}{\Gamma_A}\nonumber\\
    &- \frac{{\bf q}\cdot{\bf v}}{\Gamma_A} \frac{\langle\lambda_{i j}\rangle_0}{\langle 1\rangle_0} q_j u_i -\frac{({\bf q}\cdot{\bf v})({\bf q}\cdot{\bf E}_0) \epsilon_{\rm lat}}{e \langle 1\rangle_0 \Gamma_A} +\left(1-\frac{\Gamma_E}{\Gamma_A}\right)\frac{\langle\chi_0^{(\alpha)}\rangle_0}{\langle 1^{(\alpha)}\rangle_0},
\end{align}
where $\alpha=1,2,3,4$ and the rest of the notation is the same as in our main text.
The ``normalization condition'' reads
\begin{equation}
  \sum_{\alpha} \langle\chi_0^{(\alpha)}\rangle_0 = 0.
\end{equation}

Taking the Fermi-surface average of Eq.~(\ref{eq:1}), we have
\begin{equation}
  \frac{\Gamma_E}{\Gamma_A} \langle\chi_0^{(\alpha)}\rangle_0 = -\frac{\langle \lambda_{i j}^{(\alpha)}\rangle_0 \omega q_j u_i}{\Gamma_A} + \frac{\langle 1^{(\alpha)}\rangle_0}{\langle 1\rangle_0} \frac{\langle\lambda_{i j}\rangle_0 \omega q_j u_i}{\Gamma_A} + \frac{\langle 1^{(\alpha)}\rangle_0}{\langle 1\rangle_0} \frac{\omega \epsilon_{\rm lat}}{e}\frac{{\bf q}\cdot{\bf E}_0}{\Gamma_A}.
\end{equation}
Then,
\begin{equation}
  \label{eq:2}
  \frac{\Gamma_E}{\Gamma_A} \sum_\alpha \langle\chi_0^{(\alpha)}\rangle_0 = - \frac{\langle\lambda_{i j}\rangle_0 \omega q_j u_i}{\Gamma_A} + \frac{\langle\lambda_{i j}\rangle_0 \omega q_j u_i}{\Gamma_A} + \frac{\omega \epsilon_{\rm lat}}{e} \frac{{\bf q}\cdot{\bf E}_0}{\Gamma_A},
\end{equation}
where we have used
\begin{align}
  &\sum_\alpha \langle\lambda_{i j}^{(\alpha)}\rangle_0 = \langle\lambda_{i j}\rangle_0\nonumber\\
  &\sum_\alpha \langle 1^{(\alpha)}\rangle_0 = \langle 1\rangle_0.
\end{align}
Because of the normalization condition, Eq. (\ref{eq:2}) yields ${\bf q}\cdot{\bf E}_0 = 0$.
If we neglect the transverse part of the electric field, we have ${\bf E}_0=0$.
Accordingly,
\begin{align}
\label{eq:4nodechi0B0}
\chi_0^{(\alpha)} &= -\frac{\lambda_{i j}^{(\alpha)} \omega q_j u_i}{\Gamma_A} + \frac{\langle \lambda_{i j}\rangle_0}{\langle 1\rangle_0} \frac{\omega q_j u_i}{\Gamma_A} - \frac{m \omega^2 v_i^{(\alpha)} u_i}{\Gamma_A} 
    - \frac{{\bf q}\cdot{\bf v}}{\Gamma_A} \frac{\langle\lambda_{i j}\rangle_0}{\langle 1\rangle_0} q_j u_i +\left(1-\frac{\Gamma_E}{\Gamma_A}\right)\frac{\langle\chi_0^{(\alpha)}\rangle_0}{\langle 1^{(\alpha)}\rangle_0},
\end{align}
where
\begin{equation}
  \langle\chi_0^{(\alpha)}\rangle_0 =-\frac{\langle\lambda_{i j}^{(\alpha)}\rangle_0 \omega q_j u_i}{\Gamma_E} + \frac{\langle 1^{(\alpha)}\rangle_0}{\langle 1\rangle_0}\frac{\langle\lambda_{i j}\rangle_0 \omega q_j u_i}{\Gamma_E}.
\end{equation}

%{\bf Remark:}

Because of the time-reversal symmetry, the deformation potential and the Fermi-level density of states must be the same for time-reversed nodes.
Therefore, the 4-node model reduces to two identical copies of the 2-band model studied in our main text.
Namely, using
\begin{align}
  \langle 1\rangle_0 &= 2 \langle 1^{(1)}\rangle_0 + 2\langle 1^{(2)}\rangle_0\nonumber\\
  \langle\lambda_{i j}\rangle_0 &= 2\langle \lambda_{i j}^{(1)}\rangle_0 + 2\langle \lambda_{i j}^{(2)}\rangle_0,
\end{align}
we arrive at
\begin{align}
  \label{eq:fin}
  \langle\chi_0^{(1)}\rangle_0 &= -\frac{\omega q_j u_i}{\Gamma_E \langle 1\rangle_0} 2\left(\langle\lambda_{i j}^{(1)}\rangle_0 \langle 1^{(2)}\rangle_0 - \langle\lambda_{i j}^{(2)}\rangle_0 \langle 1^{(1)}\rangle_0 \right)\nonumber\\
  \langle\chi_0^{(3)}\rangle_0 &= \langle\chi_0^{(1)}\rangle_0 \nonumber\\
  \langle\chi_0^{(4)}\rangle_0 &= \langle\chi_0^{(2)}\rangle_0=-\langle\chi_0^{(1)}\rangle_0.
\end{align}
The first line of Eq.~(\ref{eq:fin}) is essentially identical to the expression for $\langle\chi^{(1)}\rangle_0$ in the 2-band model given by Eq.~(\ref{avchi0}).
We re-write Eq.~(\ref{eq:4nodechi0B0}) neglecting the unimportant mass term associated with $m$, 
\begin{align}\label{chi04}
\chi_0^{(\alpha)} &= -\frac{\lambda_{i j}^{(\alpha)} \omega q_j u_i}{\Gamma_A} + \frac{\langle \lambda_{i j}\rangle_0}{\langle 1\rangle_0} \frac{\omega q_j u_i}{\Gamma_A} 
    - \frac{{\bf q}\cdot{\bf v}}{\Gamma_A} \frac{\langle\lambda_{i j}\rangle_0}{\langle 1\rangle_0} q_j u_i +\left(1-\frac{\Gamma_E}{\Gamma_A}\right)\frac{\langle\chi_0^{(\alpha)}\rangle_0}{\langle 1^{(\alpha)}\rangle_0},
\end{align}

Next, we adopt a similar formalism as in Sec. (IV) to derive the solution of BKE in presence of magnetic field, such that the expression of $\chi_{1}$ for node (1) may be written as,
\begin{equation}\label{4nodechi11}
\begin{split}
\chi_{1}^{(1)} & = \frac{e}{\hbar}(\textbf{B}\cdot\boldsymbol{\Omega}_{\textbf{p}}^{(1)})\bigg(\frac{\textbf{q}\cdot\textbf{v}^{(1)}}{\Gamma_{A}}\bigg)\frac{\langle\lambda_{ij}\rangle_{0}}{\langle1\rangle_{0}}q_{j}u_{i} + i \textbf{q}\cdot\bigg(\frac{e}{\hbar}(\boldsymbol{\Omega}_{\textbf{p}}^{(1)}\cdot\textbf{v}^{(1)})\textbf{B} - \partial_{\textbf{p}}(\textbf{m}_{\bf{p}}^{(1)}\cdot\textbf{B})\bigg)\frac{iq_{j}u_{i}}{\Gamma_{A}}\frac{\langle\lambda_{ij}\rangle_{0}}{\langle1\rangle_{0}}\\
&\quad -\frac{ie\omega}{\Gamma_{A}}(\textbf{v}\times\textbf{B})\cdot \textbf{u} + \frac{i(\textbf{q}\cdot\textbf{v}-\omega)}{\Gamma_{A}}\bigg[\frac{i\textbf{q}\cdot\textbf{E}_{1}(\epsilon_{lat})}{e\langle1\rangle_{0}}\bigg]-\frac{e\textbf{v}\cdot\textbf{E}_{1}}{\Gamma_{A}}\\
&\quad +\frac{\langle\chi_{1}^{(1)}\rangle_{0}}{\langle1^{(1)}\rangle_{0}} - \frac{\Gamma_{E}}{\Gamma_{A}}\frac{\langle\chi_{1}^{(1)}\rangle_{0}}{\langle1^{(1)}\rangle_{0}} -\frac{e}{\hbar}(\textbf{B}\cdot\boldsymbol{\Omega}_{\textbf{p}}^{(1)})\bigg\{\frac{i\omega^{2}}{\Gamma_{A}^{2}}\bigg(\lambda_{ij}^{(1)} - \frac{\langle\lambda_{ij}\rangle_{0}}{\langle1\rangle_{0}}\bigg)q_{j}u_{i} + i\omega\bigg(\frac{\textbf{q}\cdot\textbf{v}}{\Gamma_{A}^{2}}\bigg)\frac{\langle\lambda_{ij}\rangle_{0}}{\langle1\rangle_{0}}q_{j}u_{i}\\
&\quad -\frac{i\omega}{\Gamma_{A}}\frac{\langle\chi_{0}^{(1)}\rangle_{0}}{\langle1^{(1)}\rangle_{0}} + i\omega\frac{\langle\chi_{0}^{(1)}\rangle_{0}}{\langle1^{(1)}\rangle_{0}}\frac{\Gamma_{E}}{\Gamma_{A}^{2}}\bigg\} + i\textbf{q}\cdot\bigg(\frac{e}{\hbar}(\boldsymbol{\Omega}_{\textbf{p}}^{(1)}\cdot\textbf{v}^{(1)})\textbf{B} - \partial_{\textbf{p}}(\textbf{m}_{\bf{p}}^{(1)}\cdot\textbf{B})\bigg)\bigg\{-\frac{\langle\chi_{0}^{(1)}\rangle_{0}}{\langle1^{(1)}\rangle_{0}}\\
&\quad \times \bigg(\frac{1}{\Gamma_{A}}-\frac{\Gamma_{E}}{\Gamma_{A}^{2}}\bigg) + \frac{\omega q_{j}u_{i}}{\Gamma_{A}^{2}}\bigg(\lambda_{ij}^{(1)} - \frac{\langle\lambda_{ij}\rangle_{0}}{\langle1\rangle_{0}}\bigg) + \frac{\textbf{q}\cdot\textbf{v}}{\Gamma_{A}^{2}}\frac{\langle\lambda_{ij}\rangle_{0}}{\langle1\rangle_{0}}q_{j}u_{i}\bigg\}.\\
&\quad 
\end{split}
\end{equation}
Similar equations as above will be written for $\chi_{1}^{(2)}$, $\chi_{1}^{(3)}$ and $\chi_{1}^{(4)}$ corresponding to nodes (2), (3) and (4) by changing the superscript (1) in above equation to $(2)$, (3) and (4), respectively.

To derive an expression of $E_{1}$, we need to take the Fermi surface average of Eq.~(\ref{4nodechi11}) around each node,
\begin{equation}\label{4nodeavchi1plus}
\begin{split}
\langle\chi_{1}^{(1)}\rangle_{0} &= \frac{e}{\hbar}\langle(\textbf{B}\cdot\boldsymbol{\Omega}_{\textbf{p}}^{(1)})(\textbf{q}\cdot\textbf{v}^{(1)})\rangle_{0}\frac{\langle\lambda_{ij}\rangle_{0}}{\langle1\rangle_{0}\Gamma_{A}}q_{j}u_{i} + i \textbf{q}\cdot\bigg(\frac{e}{\hbar}\langle(\boldsymbol{\Omega}_{\textbf{p}}^{(1)}\cdot\textbf{v}^{(1)})\rangle_{0}\textbf{B} -\langle \partial_{\textbf{p}}(\textbf{m}_{p}^{(1)}\cdot\textbf{B})\rangle_{0}\bigg)\frac{iq_{j}u_{i}}{\Gamma_{A}}\frac{\langle\lambda_{ij}\rangle_{0}}{\langle1\rangle_{0}}\\
&\quad +\frac{\omega}{\Gamma_{A}}\frac{(\textbf{q}\cdot\textbf{E}_{1})\epsilon_{lat}\langle1^{(1)}\rangle_{0}}{e\langle1\rangle_{0}} + \langle\chi_{1}^{(1)}\rangle_{0} - \frac{\Gamma_{E}}{\Gamma_{A}}\langle\chi_{1}^{(1)}\rangle_{0} -\frac{e}{\hbar}\langle(\textbf{B}\cdot\boldsymbol{\Omega}_{p}^{(1)})(\textbf{q}\cdot\textbf{v}^{(1)})\rangle_{0}\\
&\quad \times \frac{i\omega}{\Gamma_{A}^{2}}\frac{\langle\lambda_{ij}\rangle_{0}}{\langle1\rangle_{0}}q_{j}u_{i} - i\textbf{q}\cdot\bigg\{\frac{e}{\hbar}\langle(\boldsymbol{\Omega}_{p}^{(1)}\cdot\textbf{v}^{(1)})\rangle_{0}\textbf{B} - \langle \partial_{\textbf{p}}(\textbf{m}_{p}^{(1)}\cdot \textbf{B})\rangle_{0}\bigg\}\frac{\langle\chi_{0}^{(1)}\rangle_{0}}{\langle1^{(1)}\rangle_{0}}\bigg(\frac{1}{\Gamma_{A}}-\frac{\Gamma_{E}}{\Gamma_{A}^{2}}\bigg)\\
&\quad + i\textbf{q}\cdot\bigg\{\frac{e}{\hbar}\langle(\boldsymbol{\Omega}_{p}^{(1)}\cdot\textbf{v}^{(1)})\lambda_{ij}^{(1)}\rangle_{0}\textbf{B} - \langle \partial_{\textbf{p}}(\textbf{m}_{p}^{(1)}\cdot \textbf{B})\lambda_{ij}^{(1)}\rangle_{0}\bigg\}\frac{\omega q_{j}u_{i}}{\Gamma_{A}^{2}} \\
&\quad - i\textbf{q}\cdot\bigg\{\frac{e}{\hbar}\langle(\boldsymbol{\Omega}_{p}^{(1)}\cdot\textbf{v})\rangle_{0}\textbf{B} - \langle \partial_{\textbf{p}}(\textbf{m}_{p}^{(1)}\cdot \textbf{B})\rangle_{0}\bigg\}\frac{\langle\lambda_{ij}\rangle_{0}}{\langle1\rangle_{0}}\frac{\omega q_{j}u_{i}}{\Gamma_{A}^{2}}.\\
\end{split}
\end{equation}
Likewise, similar expressions as above are written for $\langle\chi_{1}^{(2)}\rangle_{0}$, $\langle\chi_{1}^{(3)}\rangle_{0}$ and $\langle\chi_{1}^{(4)}\rangle_{0}$, respectively. Imposing the normalization condition $\sum_{\alpha}\langle\chi_{1}^{(\alpha)}\rangle_{0}= 0$ and using relations $\langle\chi_{0}^{(3)}\rangle_{0} = \langle\chi_{0}^{(1)}\rangle_{0}$ and $\langle\chi_{0}^{(4)}\rangle_{0} = \langle\chi_{0}^{(2)}\rangle_{0}$, we derive the expression for $E_{1,\parallel}$ (subject to condition $\Gamma_{A}\rightarrow \infty$),
\begin{equation}\label{4nodeE1}
\begin{split}
    E_{1,\parallel} & = \bigg(\frac{e}{\omega q\epsilon_{lat}}\bigg)\bigg[\bigg\{2i\textbf{q}\cdot\bigg(\frac{e}{\hbar}\frac{\langle\boldsymbol{\Omega}_{p}^{(1)}\cdot\textbf{v}^{(1)}\rangle_{0}\textbf{B}}{\langle1^{(1)}\rangle_{0}} - \frac{\langle \partial_{\textbf{p}}\textbf{m}_{p}^{(1)}\cdot\textbf{B}\rangle_{0}}{\langle1^{(1)}\rangle_{0}}\bigg) \langle\chi_{0}^{(1)}\rangle_{0}+ 2i\textbf{q}\cdot\bigg(\frac{e}{\hbar}\frac{\langle\boldsymbol{\Omega}_{p}^{(2)}\cdot\textbf{v}^{(2)}\rangle_{0}\textbf{B}}{\langle1^{(2)}\rangle_{0}}\\
    &\quad - \frac{\langle \partial_{\textbf{p}}\textbf{m}_{p}^{(2)}\cdot\textbf{B}\rangle_{0}}{\langle1^{(2)}\rangle_{0}}\bigg)\langle\chi_{0}^{(2)}\rangle_{0}\bigg\}\bigg].\\
\end{split}
\end{equation}

We plug the expression of electric field in the Eq.~(\ref{4nodeavchi1plus}), subject to condition $\Gamma_{A}\rightarrow \infty$:

\begin{equation}\label{4nodetotalavchi}
\begin{split}
    \langle\chi_{1}^{(1)}\rangle_{0} &=\bigg\{ \frac{3e}{\hbar}\frac{\langle(\textbf{B}\cdot\boldsymbol{\Omega}_{\textbf{p}}^{(1)})(\textbf{q}\cdot\textbf{v}^{(1)})\rangle_{0}}{\Gamma_{E}}\frac{\langle\lambda_{ij}\rangle_{0}}{\langle1\rangle_{0}}q_{j}u_{i} -\frac{2e}{\hbar}\frac{\langle(\textbf{v}^{(1)}\cdot\boldsymbol{\Omega}_{\textbf{p}}^{(1)})(\textbf{B}\cdot\textbf{q})\rangle_{0}}{\Gamma_{E}}\frac{\langle\lambda_{ij}\rangle_{0}}{\langle1\rangle_{0}} q_{j}u_{i}\bigg\}\\
    &\quad- i\textbf{q}\cdot\bigg(\frac{e}{\hbar}\langle(\boldsymbol{\Omega}_{\textbf{p}}^{(1)}\cdot \textbf{v}^{(1)})\rangle_{0}\textbf{B} -\langle \partial_{\textbf{p}}(\textbf{m}_{p}^{(1)}\cdot\textbf{B})\rangle_{0}\bigg)\frac{\langle\chi_{0}^{(1)}\rangle_{0}}{\langle1^{(1)}\rangle_{0}}\bigg(\frac{1}{\Gamma_{E}}\bigg)\\
    &\quad + \frac{\omega q\epsilon_{lat}}{\Gamma_{E} e}\bigg(\frac{e}{\omega q\epsilon_{lat}}\bigg)\frac{\langle1^{(1)}\rangle_{0}}{\langle1\rangle_{0}}\bigg[\bigg\{2i\textbf{q}\cdot\bigg(\frac{e}{\hbar}\frac{\langle\boldsymbol{\Omega}_{p}^{(1)}\cdot\textbf{v}^{(1)}\rangle_{0}\textbf{B}}{\langle1^{(1)}\rangle_{0}} - \frac{\langle \partial_{\textbf{p}}\textbf{m}_{p}^{(1)}\cdot\textbf{B}\rangle_{0}}{\langle1^{(1)}\rangle_{0}}\bigg) \langle\chi_{0}^{(1)}\rangle_{0}+ 2i\textbf{q}\cdot\bigg(\frac{e}{\hbar}\frac{\langle\boldsymbol{\Omega}_{p}^{(2)}\cdot\textbf{v}^{(2)}\rangle_{0}\textbf{B}}{\langle1^{(2)}\rangle_{0}}\\
    &\quad - \frac{\langle \partial_{\textbf{p}}\textbf{m}_{p}^{(2)}\cdot\textbf{B}\rangle_{0}}{\langle1^{(2)}\rangle_{0}}\bigg)\langle\chi_{0}^{(2)}\rangle_{0}\bigg\}\bigg].
\end{split}
\end{equation}
Similar expressions are derived for $\langle\chi_{1}^{(2)}\rangle_{0}$, $\langle\chi_{1}^{(3)}\rangle_{0}$ and $\langle\chi_{1}^{(4)}\rangle_{0}$ with corresponding superscripts changed from (1) to (2), (3) and (4), respectively. The terms within the square brackets denote the electric field, they will remain the same with no changes in superscript.

For the elasticity equations in this model, we follow the formalism as in Sec.~(V) and use the relations $\langle\lambda_{hk}^{(1)}\chi_{1}^{(1)}\rangle_{0} = \langle\lambda_{hk}^{(3)}\chi_{1}^{(3)}\rangle_{0}$ and $\langle\lambda_{hk}^{(2)}\chi_{1}^{(2)}\rangle_{0} = \langle\lambda_{hk}^{(4)}\chi_{1}^{(4)}\rangle_{0}$, such that Eq.~(\ref{drag4}) modifies to,

\begin{equation}\label{4nodedrag3}
\begin{split}
    F_{h} & \simeq 2iq_{k}\bigg[\langle\lambda_{hk}^{(1)}\rangle_{0}\delta\mu_{1}+\langle\lambda_{hk}^{(2)}\rangle_{0}\delta\mu_{1} -\langle\lambda_{hk}^{(1)}\chi_{1}^{(1)}\rangle_{0}- \langle\lambda_{hk}^{(2)}\chi_{1}^{(2)}\rangle_{0}\bigg] +\text{ terms independent of $B$}.\\
\end{split}
\end{equation}
The above equation amounts essentially to Eq.~(\ref{drag4}) multiplied by two (namely, the two pairs of nodes related by time reversal contributed equally).
With the modified expressions of the drag force (Eq.~\ref{4nodedrag3}) and in the limit of $\Gamma_{A}\rightarrow\infty$, the elasticity equation given by Eq.~(\ref{eigenscalar}) changes to 
\begin{equation}\label{4nodeeigenscalar}
\begin{split}
    \rho\omega^2 u_h &=s_{hkim} q_{k}q_{m} u_{i} +\frac{e}{\hbar}\left((\textbf{B}\cdot\textbf{q})\omega u_{i}\delta_{hi} - q_{h}(\omega u_{i}B_{i})\right)\langle\langle\Omega_j(p_j - mv_j)f_0\rangle\rangle_0\\
  &-2iq_{k}\delta\mu_1\langle\lambda_{hk}\rangle_0 + 2i q_k \langle\lambda_{hk}\chi_{1}\rangle_0.\\
\end{split}
\end{equation}

In this model, for the limit $\Gamma_{A}\rightarrow \infty$, the transverse modes corresponding to Eqs.~(\ref{Mxyz}) and (\ref{Gtransxy}) do not exhibit the PMC effect. Hence, we concentrate on the dispersion of the longitudinal mode corresponding to Eq.~(\ref{4nodeeigenscalar}). In this case, Eq.~(\ref{2nodeft}) changes to

\begin{equation}\label{4nodeft}
    2\bigg\{iq_{z}\langle\lambda_{zz}^{(1)}\rangle_{0}\delta\mu_{1} + iq_{z}\langle\lambda_{zz}^{(2)}\rangle_{0}\delta\mu_{1}\bigg\} \simeq -2q_{z}^{2} \frac{\langle\lambda_{zz}\rangle_{0}}{\langle1\rangle_{0}}\frac{\epsilon_{lat}}{e} E_{1,\parallel}.
\end{equation}
In calculating the above expression, we plug the value of electric field $E_{1,\parallel}$ from Eq.~(\ref{4nodeE1}) and use the expressions from Eq.~(\ref{eq:fin}) and Eq.~(\ref{eq:relations}), such that
\begin{equation}\label{4finalinte2}
\begin{split}
    2\bigg\{iq_{z}\langle\lambda_{zz}^{(1)}\rangle_{0} + iq_{z}\langle\lambda_{zz}^{(2)}\rangle_{0}\bigg\}\delta\mu_{1} \simeq  4iq_{z}^{2} q_{z}u_{z}\frac{\langle\lambda_{zz}\rangle_{0}}{\langle1\rangle_{0}}\bigg[\frac{eB_{z}|C|}{3\pi^{2}\hbar^{2}}\frac{1}{\Gamma_{E}}\bigg(\frac{\langle\lambda_{zz}^{(1)}\rangle_{0}}{\langle1^{(1)}\rangle_{0}}-\frac{\langle\lambda_{zz}^{(2)}\rangle_{0}}{\langle1^{(2)}\rangle_{0}}\bigg)\bigg].\\
\end{split}
\end{equation}

Invoking similar approximations as $\Gamma_{A}\rightarrow \infty$ and using Eqs.~(\ref{4nodechi11}) and (\ref{4nodetotalavchi}), we find 
\begin{equation}\label{4nodeinte3}
\begin{split}
-2\bigg\{ iq_{z}\langle\lambda_{zz}^{(1)}\chi_{1}^{(1)}\rangle_{0} + iq_{z}\langle\lambda_{zz}^{(2)}\chi_{1}^{(2)}\rangle_{0}\bigg\} \simeq \frac{2iq_{z}^{2}q_{z} u_{z}}{\Gamma_{E}}\frac{\langle\lambda_{zz}\rangle_{0}}{\langle1\rangle_{0}}\frac{eB_{z}|C|}{4\pi^{2}\hbar^{2}}\bigg(\frac{\langle\lambda_{zz}^{(1)}\rangle_{0}}{\langle1^{(1)}\rangle_{0}}- \frac{\langle\lambda_{zz}^{(2)}\rangle_{0}}{\langle1^{(2)}\rangle}\bigg) +2q_{z}^{2}q_{z}\frac{eB_{z}|C|}{\Gamma_{E}^{2}3\pi^{2}\hbar^{2}}\omega u_{z}\bigg(\frac{\langle\lambda_{zz}^{(1)}\rangle_{0}}{\langle1^{(1)}\rangle_{0}}\\
    - \frac{\langle\lambda_{zz}^{(2)}\rangle_{0}}{\langle1^{(2)}\rangle_{0}}\bigg)\bigg\{\frac{2\langle\lambda_{zz}^{(1)}\rangle_{0}}{\langle1^{(1)}\rangle_{0}}\frac{\langle1^{(2)}\rangle_{0}}{\langle1\rangle_{0}}+\frac{2\langle\lambda_{zz}^{(2)}\rangle_{0}}{\langle1^{(2)}\rangle_{0}}\frac{\langle1^{(1)}\rangle_{0}}{\langle1\rangle_{0}} - \frac{\langle\lambda_{zz}\rangle_{0}}{\langle1\rangle_{0}}\bigg\}.\\
\end{split}
\end{equation}

\begin{figure}
   \centering
   \includegraphics{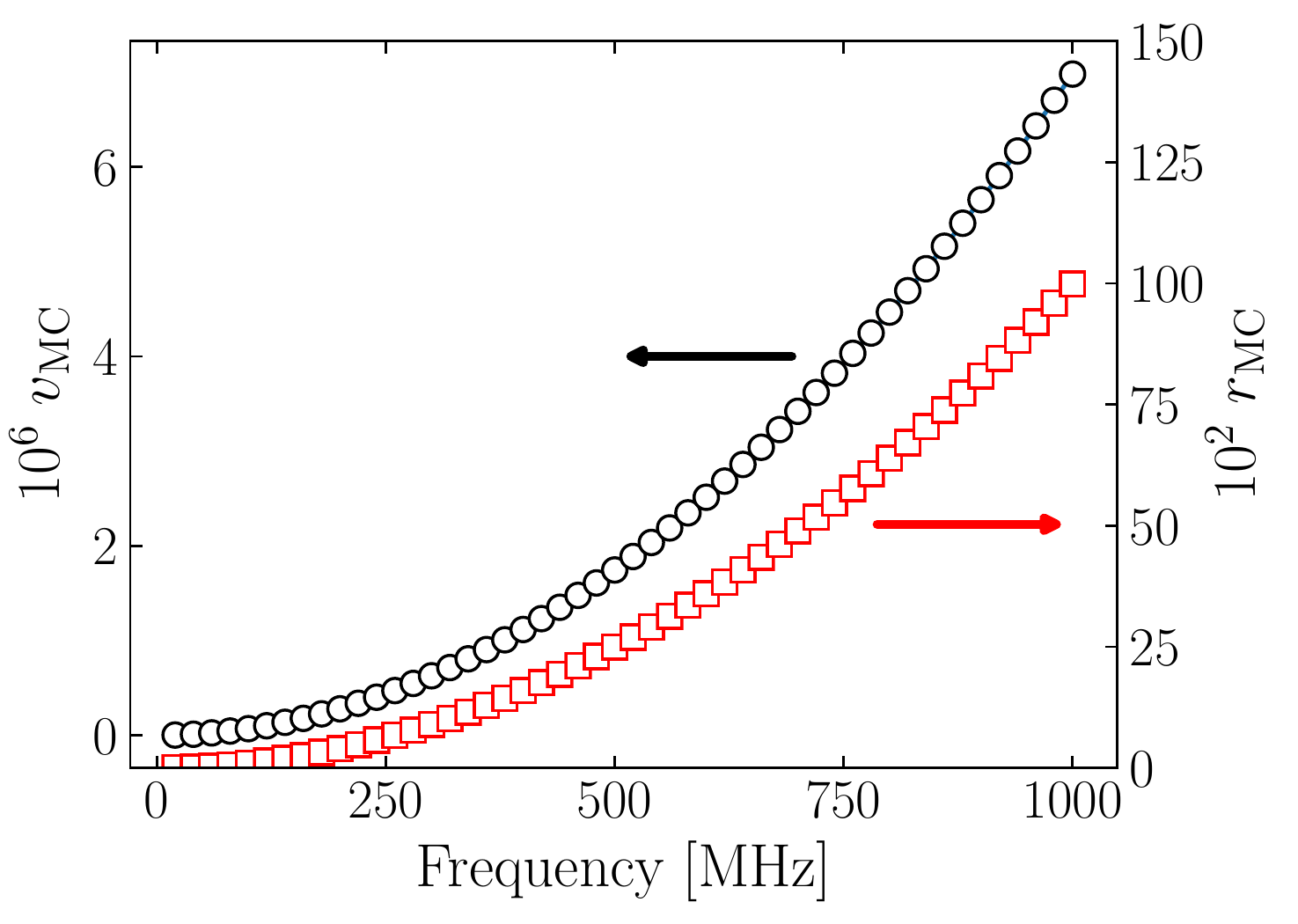}
    \caption{PMCE in a $2n$-node model. For $n=2$, the values of $v_{\rm{MC}}$ and $r_{\rm{MC}}$ are larger in magnitude than the corresponding two-node model considered in the main script, and subsequently increases with the increase in number of nodes. }
    \label{fig:pmche4node}
\end{figure}

Using Eq.~(\ref{4finalinte2}) and Eq.~(\ref{4nodeinte3}), we write the longitudinal mode corresponding to Eq.~(\ref{4nodeeigenscalar}) as
\begin{equation}\label{4nodelongmode}
\begin{split}
    0 = q_{z}^{2}s_{33} -\rho\omega^{2} -\bigg[11iq_{z}^{2} q_{z}\frac{\langle\lambda_{zz}\rangle_{0}}{\langle1\rangle_{0}}\bigg\{\frac{eB_{z}|C|}{6\pi^{2}\hbar^{2}}\frac{1}{\Gamma_{E}}\bigg(\frac{\langle\lambda_{zz}^{(1)}\rangle_{0}}{\langle1^{(1)}\rangle_{0}}-\frac{\langle\lambda_{zz}^{(2)}\rangle_{0}}{\langle1^{(2)}\rangle_{0}}\bigg)\bigg\} + 2q_{z}^{2}q_{z}\frac{eB_{z}|C|}{\Gamma_{E}^{2}3\pi^{2}\hbar^{2}}\omega \bigg\{\frac{\langle 1^{(1)}\rangle_0-\langle 1^{(2)}\rangle_0}{\langle 1^{(1)}\rangle_0+\langle 1^{(2)}\rangle_0}\\
    \times\bigg(\frac{\langle\lambda^{(1)}_{zz}\rangle_0}{\langle1^{(1)}\rangle_0}-\frac{\langle\lambda^{(2)}_{zz}\rangle_0}{\langle1^{(2)}\rangle_0}\bigg)^2\bigg\}\bigg].\\
\end{split}
\end{equation}

Solving the above equation, we find the real part of the frequency as
\begin{equation}\label{4nodewr}
\delta\omega_{R} \simeq \frac{q_{z}^{3}eB_{z}|C|}{3\pi^{2}\hbar^{2}\Gamma_{E}^{2}\rho}\bigg\{\frac{\langle 1^{(1)}\rangle_0-\langle 1^{(2)}\rangle_0}{\langle 1^{(1)}\rangle_0+\langle 1^{(2)}\rangle_0} \bigg(\frac{\langle\lambda^{(1)}_{zz}\rangle_0}{\langle1^{(1)}\rangle_0}-\frac{\langle\lambda^{(2)}_{zz}\rangle_0}{\langle1^{(2)}\rangle_0}\bigg)^2\bigg\}.\\
\end{equation}
We note that the above expression is two times the right hand side of the second line in Eq.~(\ref{eq:solong}), which is justified as we are considering a model of $2n$ nodes with $n=2$. The correction to the velocity of sound is then

\begin{equation}\label{4nodevelocity}
\begin{split}
\delta c_{s} &\simeq \frac{\partial \delta\omega_{R}}{\partial |q_{z}|}
\simeq q_z |q_z| \frac{e B_z |C|}{\pi^2 \hbar^2\rho}\frac{1}{\Gamma_E^2}\frac{\langle 1^{(1)}\rangle_0-\langle 1^{(2)}\rangle_0}{\langle 1^{(1)}\rangle_0+\langle 1^{(2)}\rangle_0} \bigg(\lambda_1^{(1)}-\lambda_1^{(2)} + \frac{\lambda_2^{(1)}}{3}-\frac{\lambda_2^{(2)}}{3}\bigg)^2,
\end{split}
\end{equation}
where we have used the relation given by Eq.~(\ref{relation2}). Therefore, the phonon magnetochiral effect in the sound velocity is given as
\begin{equation}\label{4nodevmc}
\begin{split}
v_{\rm{MC}}&\simeq \frac{c_{s}(\textbf{B}\parallel\hat{\textbf{q}})-c_{s}(\textbf{B}\parallel-\hat{\textbf{q}})}{c_{s}(0)}\\
& \simeq \frac{2e q_{z} |q_z|B_{z}|C|}{\pi^2\hbar^2\rho c_{s}(0)}\frac{1}{\Gamma_{E}^{2}}\frac{\langle1^{(1)}\rangle_0 - \langle1^{(2)}\rangle_0}{\langle1^{(1)}\rangle_{0}+\langle1^{(2)}\rangle_{0}}\bigg(\lambda_1^{(1)}-\lambda_1^{(2)} + \frac{\lambda_2^{(1)}}{3}-\frac{\lambda_2^{(2)}}{3}\bigg)^{2}.
\end{split}
\end{equation}
Once again, we note the PMCE in the velocity for the model with $n=2$, is enhanced by a factor of 2 compared to the PMCE for the two-node model. Thus for a model of $2n$ nodes, the increment in the PMCE in velocity will be $n$ times greater than that of the simple two-node model.

Similarly from Eq.~(\ref{longmode}), the imaginary part of the frequency is given as
\begin{equation}\label{4nodewi}
\delta\omega_{I}(\textbf{q},\textbf{B}) \approx  -\bigg[\frac{11}{12}q_{z}| q_{z}|\frac{\langle\lambda_{zz}\rangle_{0}}{\langle1\rangle_{0}}\bigg\{\frac{eB_{z}|C|}{\pi^{2}\hbar^{2}c_{s}(0)\rho_{d}}\frac{1}{\Gamma_{E}}\bigg(\frac{\langle\lambda_{zz}^{(1)}\rangle_{0}}{\langle1^{(1)}\rangle_{0}}-\frac{\langle\lambda_{zz}^{(2)}\rangle_{0}}{\langle1^{(2)}\rangle_{0}}\bigg)\bigg\}\bigg].
\end{equation}
Using the above equation, we write the attenuation of the sound, quantified by factor $r_{\rm{MC}}$ given by Eq.~(\ref{rmc}).
Considering similar material parameters as used for the two-node model, we present the numerical estimates of PMCE in the velocity and attenuation of sound waves in Fig.~\ref{fig:pmche4node} for the four-node model. As can be seen, the results are qualitatively similar to the 2-node model, but are quantitatively enhanced. For the $2n$-nodes model, with $n>1$, we expect the PMCE will be higher compared to the two node model and the effect will subsequently increase with the number of nodes, $n$.

\subsection*{VIII. Phonon magnetochiral effect (PMCE) in a model of multifold chiral fermions}

The formalism developed in the previous subsections for a model of WSM with 2$n$ nodes ($n\geq 1$) may be extended to address recently discovered chiral crystals that exhibit multifold fermions. 
The materials in this group are represented by ASi, AGe (A = Rh, Co) and AlPt. 
Their band dispersions host a spin$-3/2$ fermion at the $\Gamma$-point, and a double spin-1 fermion at the $R$ point \cite{tang2017,dejuan2017}. 
Depending on the position of the Fermi level, two cases have been predicted: (i) the $\Gamma$-point exhibits a hole Fermi pocket whereas the $R$-point hosts an electron Fermi pocket \cite{tang2017}; (ii) both the $\Gamma$-point and the $R$-point host an electron Fermi pocket \cite{dejuan2017}. 
In case of (i), the two bands crossing the Fermi energy near the $\Gamma$-point have Chern numbers $C =3$ and $C=1$.
In case of (ii), the two bands crossing the Fermi energy near the $\Gamma$-point have Chern numbers $C =-3$ and $C=-1$.
In both cases, the bands crossing the Fermi energy near the $R$-point have Chern numbers $C =-2,-2, 0, 0$. 
Interestingly, the bands with $C=0$ have nonzero orbital magnetic moment  \cite{dejuan2017}.

The aforementioned Fermi surfaces are more complicated than the ones assumed in our model. 
First, for each given chirality, there are multiple bands that cross the Fermi surface, each of which can in general have a different Chern number and a different Fermi velocity.
Second, while our assumption of spherically symmetric Fermi surfaces (adopted for simplicity) may describe well the Fermi pocket surrounding the $\Gamma$ point, departures from spherical symmetry may be significant at the Fermi pocket enclosing the $R$ point.  
Indeed, a full quantitative theory of the phonon magnetochiral effect in these materials requires a multiband Boltzmann equation, with multiple scattering rates and anisotropic Fermi surfaces.
While carrying out such theory lies outside the scope of the present work, we will make qualitative statements as to how our results are expected to change in more realistic settings.

First, we will map the electronic bands near the $\Gamma$-point to a single, linearly dispersing isotropic band with Chern number $C=+4$ and $C=-4$ for cases (i) and (ii), respectively. 
This mapping to a single effective band may be qualitatively reasonable if the scattering rate connecting the two bands crossing the Fermi level is strong enough.
In a similar fashion, the two nearly-degenerate $C=-2$ bands emanating from the $R$-point may be qualitatively modeled as a single, isotropic and linearly dispersing band with $C=-4$.
Then, we may reuse Eqs.~(\ref{eo})-(\ref{gradm}) for the dispersion at $\Gamma$ and $R$. 
For now, we will ignore the $C=0$ bands in the vicinity of the $R$ point.
% and will comment on their contribution below.

If the $\Gamma$-point exhibits a hole Fermi pocket, the overall form of our result for the PMCE in the sound velocity (given by Eq.~(\ref{vmc})) and sound attenuation (given by Eq.~(\ref{rmc})) will still apply, except for the fact that $|C| = 4$. Thus, we expect that the PMCE will be {\em enhanced} compared to the two-node Weyl semimetal model.

If the $\Gamma$-point exhibits an electron Fermi pocket, the single effective band near $\Gamma$ will have a Chern number of $-4$, much like the single effective band near $R$. 
In our $2-$node model, this situation can be captured by requiring that one node is "electron-doped" and the other is  "hole-doped". 
In such situation, the Berry curvatures at the Fermi surface are identical for the two nodes, {\em but} the Fermi velocities have opposite signs.
Because the product of the Berry curvature and the electronic velocity changes sign from one node to another, the result for the PMCE (Eqs.~(\ref{vmc}) and (\ref{rmc})) will remain qualitatively unchanged with respect to the case in which the $\Gamma$ point exhibits a hole Fermi pocket.

Thus far, we have ignored the $C=0$ bands in the vicinity of the $R$ point.
Having a nonzero orbital magnetic moment, they too will contribute to the PMCE. We expect that this contribution will be of similar form and similar order of magnitude to the aforementioned contribution from the $C=-2$ bands.

%-----------------------------------

\end{document}